%% file: Final.tex
\documentclass[usenatbib,]{mnras}
\usepackage{graphicx}	
\usepackage{amsmath}	
\usepackage{amssymb}	
\usepackage{multicol}        
\usepackage{bm}		
\usepackage{pdflscape}	

\usepackage[T1]{fontenc}
\usepackage{ae,aecompl}
\usepackage{xcolor}
\usepackage{soul}
\usepackage{ulem}

\usepackage{newtxtext,newtxmath}

\newcommand{\kms}{\,km\,s$^{-1}$} 

\input definitions

\title[The physics of the K-S relation] {Gravity or turbulence? VI. The physics behind the Kennicutt-Schmidt relations}
\author[Javier Ballesteros-Paredes et al.]{Javier Ballesteros-Paredes$^{1}$, 
Manuel Zamora-Avil\'es$^2$, 
Carlos Román-Z\'u\~niga$^3$, 
Aina Palau$^1$, 
\newauthor
Bernardo Cervantes-Sodi$^1$,
Karla Guti\'errez-D\'avila$^1$,
Vianey Camacho$^1$, 
Eric Jim\'enez-Andrade$^1$, 
\newauthor and Adriana Gazol$^1$
\thanks{e-mail: j.ballesteros@irya.unam.mx}%
\\ 
\\
$^{1}$Universidad Nacional Aut\'onoma de M\'exico, Instituto de Radioastronom\'ia y Astrof\'isica.\\
Antigua Carretera a P\'atzcuaro 8701, Ex-Hda. San Jos\'e de la Huerta, 58089, Morelia, Michoac\'an, M\'exico
\\
$^{2}$INAOE
\\
$^{3}$Universidad Nacional Aut\'onoma de M\'exico, Instituto de Astronom\'ia, campus Ensenada. 
}
\date{Last updated 2024, June 8th; in original form 2023, September 30th}

\pubyear{2024}

\begin{document}
\label{firstpage}
\pagerange{\pageref{firstpage}--\pageref{lastpage}}
\maketitle

\begin{abstract}

{ We explain the large variety of star formation laws in terms of one single, simple law that can be inferred from the definition of the star formation rate and basic algebra. The resulting equation, $\SFR = \eff\ \Mcollapsing/\tauff$, although it has been presented elsewhere, is interpreted in terms of clouds undergoing collapse { rather than being turbulence-supported, an idea that different groups have pursued this century}. Under such assumption, one can explain the constancy of $\eff$, the different intra-cloud correlations observed in Milky Way's molecular clouds, as well as the resolved and unresolved extragalactic relationships between SFR and a measurement of the mass in CO, HCN, and CO+HI. We also explain why the slope of the correlation changes when the orbital time $\tauorb$ is considered instead of the free-fall time, and why estimations of the free-fall time from extragalactic observations skew the correlation, providing a false sublinear correlation. We furthermore show that the apparent nearly linear correlation between the star formation rate and the dynamical equilibrium pressure in the midplane of the galaxies, $\PDE$, is just a consequence of $\PDE$ values being dominated by the variation of the column density of molecular gas. All in all, we argue that the star formation law is driven by the collapse of cold, dense gas, which happens to be primarily molecular in the present Universe, and that the role of stellar feedback is just to shut down the star formation process, not to shape the star formation law.}
\end{abstract}

\begin{keywords}
ISM: clouds - stars: formation - galaxies: general - ISM: kinematics and dynamics.
\end{keywords}



\section{Introduction}\label{sec:intro}

More than 60 years ago \citet{Schmidt59} argued that ``it would seem most probable that the rate of star formation depends on [...] a power of the gas density'', { that is},
\begin{equation}
  \SFR \propto \rhogas^M  ,
  \label{eq:conjetura}
\end{equation}
{ For many years, different authors intended to correlate the star formation rate and (some measurement of) the gas density. As a result, a variety of results were reported, with substantial scatter among them \citep{Sanduleak69, Madore+74, Hamajima_Tosa75, Berkhuijsen77,  Talbot80, Kennicutt89, Buat+89}. Consequently, it was not clear whether there was a single law, different ones or none at all.}  Finally, almost 30 years later, \citet{Kennicutt98} published a correlation in terms of the surface densities of star formation rate and gas mass, $\Sigmasfr$ and $\Sigmagas$, respectively. \citet{Kennicutt98} showed that the column density of star formation rate in galaxies scales with the column density of gas as
\begin{equation}
  \Sigmasfr \propto \Sigmagas^N  .
  \label{eq:KS_columnar}
\end{equation}

The power-law index found by \citet{Kennicutt98} was $N\sim1.4$. Since then, there has been an enormous amount of work exploring different environments and scales, from local molecular clouds to ultraluminous infrared galaxies \citep[see, e.g., ][for a review]{Kennicutt_Evans12}, and each environment and scale seems to provide a different value for $N$, from 1 to near 4. 

{ There is a variety of interpretations trying to explain the different tendencies observed in the literature. For instance some} authors have tried to explain the \citet{Kennicutt98} correlation in terms of a nearly constant width $h$ of the galactic disk, such that if $\Sigmasfr\propto \Sigmagas/\tauff$, and $\tauff\propto \rho^{-1/2} \propto (\Sigmagas/h)^{-1/2}$, then $\Sigmasfr\propto\Sigmagas^{3/2}$ \citep[e.g., ][]{Elmegreen02}. Others have tried to explain the correlation in terms of whether the galactic disk can gather material in timescales comparable to the orbital time and produce star formation \citep[e.g., ][]{Kennicutt89, Tan00, Li+05, Bonnell+13}. Another group of authors have tried to explain the observed correlation in terms of turbulence providing support to clouds against collapse \citep[e.g., ][]{Krumholz_McKee05, Padoan+12, Padoan+16b, Kim+21, Evans+22}, while others tried modelling the collapse of a cloud that is accreting gas from large-scale colliding flows \citep{Zamora-Aviles+12, Zamora-Aviles_Vazquez-Semadeni14}. { In addition, it has been acknowledged that the relationship between the column densities of star formation and gas is better suited if the latter is divided by the estimated free-fall time of the gas \citep[e.g.,][]{Krumholz+12, Elmegreen18}, { and that this occurs in regions which can rapidly cool down to become dense and collapse} within a free-fall time \citep[][]{Clark_Glover14}. } 

{ In terms of observations of external galaxies, there are two types of results: (a) when total gas column density (atomic + molecular) is measured, the correlation between $\Sigmasfr$ and $\Sigmagas$ is superlinear. (b) When only molecular gas is observed, it tends to be shallower, typically close to linear. (c) Both relations are shallower if the corresponding gas column density (total or molecular) is divided by a timescale (orbital or free-fall). This occurs either for whole galaxies \citep[e.g., ][]{Kennicutt98, Daddi+10, Krumholz+12, Kennicutt_DeLosReyes21}, as well as for resolved galaxies \citep[e.g., ][]{Bigiel+08, Sun+23, Jimenez-Donaire+23, Elmegreen24, Ellison+24}. }

{ { More specifically, in the case of whole galaxies, } \citet{Kennicutt_DeLosReyes21} confirm that the former correlation found by \citet{Kennicutt98} for unresolved galaxies using total gas (molecular and atomic) stands for a sample of starburst and non-starburst galaxies with a power-law index of $N\sim$~1.5. However, each sample exhibits different tendencies, with the starburst galaxies having indexes closer to 1 at larger column densities of gas mass and star formation. In contrast, non-starburst galaxies exhibit indexes of about 1.34, consistent with the 1.4 found originally by \citet{Kennicutt98}. 

In the case of resolved galaxies, \citet{Bigiel+08} } have found, for a few galaxies resolved at scales of $\sim$1~kpc, a linear correlation between $\Sigmasfr$ { and the column density of molecular gas, $\SigmaMC$. However, the power-law index of the correlation increases to 1.5--2.7 when the total column density (atomic and molecular gas) is used.} 

{ That work has been extended recently to 80 galaxies by \citet{Sun+23}, who confirmed the linear correlation between the column densities of star formation and molecular gas. In addition, these authors showed flatter correlations} between $\Sigmasfr$ and the ratio $\SigmaMC/\tau$, with $\tau$ either the orbital time of the galaxy or { (an estimation of\footnote{ We explicitly write ``an estimation of the free-fall time'' because we believe that this is a wrong estimation of the actual free-fall time, as we will discuss in \S\ref{sec:SFR_Sigma_over_tauff}.})} the free-fall timescale. The power-law indexes were $\sim$0.6 and 0.8, respectively. At sub-kpc scales, nonetheless, the correlation is substantially less crear \citep{Schruba+10, Williams+18, Pessa+22, Pan+22}. 

Although most of the observational studies have been extragalactic, the correlation has also been studied in Milky Way clouds \citep[see, e.g., ][and references therein]{Kennicutt_Evans12}. Since for local clouds we have substantially more detailed information than in the extragalactic studies, people have studied the star-formation activity in two different ways: (a) cloud-to-cloud correlations, where correlations similar to eq.~(\ref{eq:KS_columnar}) for a set of clouds are analyzed \citep[a case that we will call the inter-cloud case, see][]{Evans+09, Heiderman+10, Lada+10, Lada+12, Pokhrel+21, Spilker+21}; and (b) correlations for single clouds at different level \citep[e.g., ][a case that we will call the intra-cloud case]{Gutermuth+11, Lada+13, Spilker+21, Pokhrel+21}.

The main results in each case are the following: For the inter-cloud case, \citet{Evans+09, Heiderman+10} and \citet{Lada+13} found that the star formation rates in the Solar Neighborhood are substantially larger than the extragalactic estimates. In addition, these authors report a poor correlation \citep{Evans+09, Heiderman+10} or no correlation at all \citep{Lada+10} between the surface densities of star formation and gas mass for different clouds. Indeed, local clouds exhibit various star formation rates, from the Orion molecular cloud, with an active star formation ongoing, down to clouds like the Musca molecular cloud, with almost no star formation. Nonetheless, such difference was found for clouds at the same column densities, $\sim$\diezala{21}\,cm\alamenos2, implying no correlation between different clouds \citep{Lada+10}.  

Interestingly, despite the lack of a $\Sigmasfr-\Sigmagas$ correlation for local clouds, a linear correlation between the mass of \textit{dense gas} in the clouds ($\Mdense$) and the number of young stellar objects ($\Nysomath$) for the inter-cloud case \citep{Lada+10} has been observed,
\begin{equation} 
  \Nysomath\propto \Mdense  . 
  \label{eq:LadaCorrelation}
\end{equation}

To complicate the situation further, three more correlations between the surface densities of star formation rate and molecular cloud mass ($\Sigmasfr$ and $\Sigmagas$ respectively) have been reported in the case of the intra-cloud studies. On the one hand, a power-law index of $N~\sim 2$ in eq.~(\ref{eq:KS_columnar}),
\begin{equation}
  \Sigmasfr \propto \Sigmagas^2
  \label{eq:KS-Pohkrel}
\end{equation}
is frequently reported for dense clouds as Orion A, B, Monoceros R2, Ophiuchus, Perseus, etc. \citep{Gutermuth+11, Lada+13, Willis+15, Pokhrel+21}, although with some scatter in some cases. A similar correlation, but with a power-law exponent of $\sim$3, i.e., 
\begin{equation}
  \Sigmasfr \propto \Sigmagas^3
  \label{eq:KS-Lada17}
\end{equation}
was found for the California cloud by \citet{Lada+17}.  Finally, a linear correlation between $\Sigmasfr$ and the ratio $\Sigmagas/\tauff$ i.e., 
\begin{equation}
  \Sigmasfr \propto \frac{\Sigmagas}{\tauff} .
  \label{eq:Pokhrel_Sigmagas_tauff}
\end{equation}
has been found by \citet[e.g.,][]{Pokhrel+21} for most of the nearby GMCs. As a summary, Table~\ref{tab:example_table} lists the most common values of the correlations related to the star formation rate\footnote{ {Since the literature is bast, this list is necessarily incomplete. We point out only to some of the main works driving these results.}} \citep[for a compilation of values in nearby galaxies, see][]{Sun+23}.

\begin{table}
  \centering
  \caption{Types of correlations found in the literature. Power-law values are approximated since, typically, there is a large scatter. References: (1) \citet{Lada+10}, (2) \citet{Pokhrel+21} (3), \citet{Lombardi+14}, (4) \citet{Zari+16}, (5) \citet{Lada+17}, (6) \citet{Lada+13}, (7) \citet{Gao-Solomon04}, (8) \citet{Kennicutt98}, (9) \citet{Bigiel+08}, (10) \citet{Sun+23}, (11) \citet{Jimenez-Donaire+23}.  }
	\label{tab:example_table}
	\begin{tabular}{lll} 
\hline
Correlation  & Type	& References	\\
\hline
$\Nysomath\propto \Mdense$	& Galactic	& 1	\\
$\Sigmasfr \propto \SigmaMC/\tauff	$ & Galactic	& 2	\\
$\Sigmasfr \propto \SigmaMC^{2}$	& Galactic	& 2, 3, 4 \\
$\Sigmasfr \propto \SigmaMC^{3.3}$	& Galactic	& 5, 6 \\
$\SFR\propto \Mdense $	& Extragalactic	& 7	\\
$\Sigmasfr \propto \Sigmagas^{1.4}$ & Extragalactic	& 8 \\
$\Sigmasfr \propto \SigmaMC	$ & Extragalactic	& 9, 10, 11	\\
$\Sigmasfr \propto (\SigmaMC/\tauff)^{2/3}	$ & Extragalactic	& 10	\\
$\Sigmasfr \propto (\SigmaMC/\tauorb)^{4/5}	$ & Extragalactic	& 9	\\
		\hline
	\end{tabular}
\end{table}

In an attempt to understand all these different results with a single explanation, in the present work {, we adopt a similar approach as \citet{Elmegreen18}, aimed at understanding the different correlations under the light of a single, fundamental equation. The different correlations are thus explained in terms of the observational effects that the different methods may have. Thus, we first provide a simple algebraic derivation of what we  \citep[and others too, see e.g.,][]{Elmegreen18, Pokhrel+21} will call the ``fundamental law for star formation,''} which relates the star formation rate, the mass of the gas, and a characteristic timescale (\S\ref{sec:fundamental}). This type of equation has been introduced heuristically in different forms by different authors \citep[e.g., ][]{Silk97, Elmegreen97, Kennicutt98, Tan00, Elmegreen02, Krumholz_McKee05, Genzel+10, Krumholz+12, Elmegreen18, Bacchini+19a, Bacchini+19b, Pokhrel+21}, { but in the present work we derive it from the} pure definition of star formation rate, with no additional assumptions but simple algebra. With this definition, we furthermore show (\S\ref{sec:eff_cst2}) that if the time involved in the fundamental law is the free-fall timescale, and if a cloud exhibits a linear behaviour between its $\Sigmasfr$  and the ratio $\Sigmagas/\tauff$ at different levels of its column density, then the gas under consideration in the observations {\it must} be collapsing. 

{ In a companion contribution \citep{Zamora-Aviles+24}, we furthermore show that, indeed, collapsing clouds in numerical simulations follow the fundamental law and, furthermore, we show that measurements of the wrongly named ``efficiency per free-fall time'' made mimicking the observational works, found typical values $\eff\sim$~0.03, even though these clouds exhibit final efficiencies of 10\%\ after 1--2 free-fall times. Our results contrast with the common hypothesis that turbulence regulates the star formation rate and, thus, defines the star formation law. Instead, we propose that the star formation law is defined by collapse, and the final efficiency is set by the disruption of collapsing clouds due to the stellar feedback.} 

In \S\ref{sec:galactic} and \ref{sec:extragalactic}, we use the fundamental equation of star formation to derive the variety of correlations observed in local clouds and external galaxies, respectively, showing that all of them can be deduced from this equation, provided there is an understanding of the actual structure of the objects under study. In \S\ref{sec:discussion}, we discuss the implications of our results in terms of gravity being responsible for the star formation law rather than turbulence, galactic rotation, { midplane pressure, or other physical agents that could regulate the star formation process. In particular, we discuss the existence of clouds with a variety of virial parameters and its relation to the SF law, how the inner structure of clouds defines the constancy of $\eff$, why this factor cannot be considered as an efficiency per free-fall time, and how our work compares to previous work in this area.  Finally, in \S\ref{sec:conclusions}, we provide our main conclusions. 
}

\section{The fundamental equation of star formation}\label{sec:fundamental}

As summarized by \citet{Genzel+10}, a typical approach to explain the Kennicutt-Schmidt relation is to assume that the volume density of star formation rate, $\rhosfr$, scales linearly with the ratio between the volume density of the gas, $\rhogas$ and the free-fall time, $\tauff$ \citep{Schmidt59, Kennicutt98, Krumholz_McKee05}. Furthermore, noticing that 
\begin{equation}
    \tauff = \bigg(\frac{3\pi}{32 G \rho}\bigg)^{1/2}
    \label{eq:tauff}
\end{equation}
(with $G$ the universal gravitational constant and $\rho$ the mass density), one could postulate 
\begin{equation}
  \rhosfr \propto \rhogas^{3/2}.
  \label{eq:genzel}
\end{equation}
There are { two} problems with this approach. On the one hand, the relationship between $\rhosfr$ and the ratio $\rhogas/\tauff$ is a postulate, but there is no formal proof. { On the other hand,} observations show a correlation between column densities, not volume densities. Thus, eq.~(\ref{eq:genzel}) is valid simultaneously with eq.~(\ref{eq:KS_columnar}) for $N\sim3/2$ only if all the observed regions have the same depth along the line of sight. Although, in principle, one can accept that galaxy disks have similar scale heights, it is not necessarily true, provided their large dynamical range of masses, sizes, and surface brightness.

\subsection{Derivation}

By definition, the star formation rate is the time derivative of the mass of newborn stars, 
\begin{equation}
  \mathrm{SFR}=\frac{dM_*}{dt}.
  \label{eq:SFR_definition}
\end{equation}
Given the impossibility of estimating the instantaneous star formation rate observationally, astronomers have  estimated the mean star formation rate by averaging over a finite arbitrary time interval $\Delta t$, 
\begin{equation}
  \promedio{\mathrm{SFR}}_{\Delta t} = \frac{1}{\Delta t}\int_{\Delta t} \frac{dM_*}{dt} dt = \frac{M_{*,\Delta t}}{\Delta t} 
  \label{eq:meanSFR}
\end{equation}
This general equation can be applied to any timescale $\Delta t$. In particular, 
some authors have assumed that there is a characteristic timescale for star formation, { $\tausf$,  depending on the young stellar objects (YSOs) under scrutiny used to estimate the star formation rate\footnote{ In the estimations of the star formation rate for local molecular cloud studies, $\tauyso\sim2$~Myr \citep{Evans+09, Heiderman+10, Lada+10} when Class 0, I and II YSOs are considered, but $\tauyso\sim0.3$--0.5~Myr if only Class 0 and I YSOs are involved in the scrutiny \citep[e.g.,][]{Lada+13, Lombardi+14, Lada+17, Pokhrel+21}.},}  such that one can estimate the star formation rate as  
\begin{equation}
  \promedio{\mathrm{SFR}}_\tausf = \frac{\Mtotstarstausf}{\tausf} ,
\end{equation}
where we have explicitly written the subscript $\tausf$ to indicate that these quantities, and the mass converted into stars, have been computed by integrating over the characteristic star formation timescale.

{
Since we want to relate this quantity to the mass in gas, and since the units of the star formation rate are mass per unit of time, we now multiply and divide this equation by the mass in gas, $\Mgas$, and a characteristic timescale of the gas, $\tau$. Thus, defining $\epsilon$ as the ratio between this time and the depletion time, 
\begin{equation}
  \epsilon \equiv \frac{\tau} {\taudepl},
\end{equation}
where the depletion time is the time to exhaust the available mass $\Mgas$ by forming stars at the rate $\Mtotstarstausf/\tausf$, and is given by
\begin{equation}
  \taudepl = \frac{\Mgas}{\Mtotstarstausf/\tausf} ,
\end{equation}
we obtain,
\begin{equation}
  \promedio{\mathrm{SFR}}_\tausf = \epsilon\ \bigg(\frac{\Mgas}{\tau}\bigg)  .
  \label{eq:fundamental}
\end{equation}
{ This equation is not new. It has already been proposed heuristically in previous works \citep[e.g.,][]{Silk97, Elmegreen97, Tan00, Elmegreen02, Genzel+10, Krumholz+12, Elmegreen18, Pokhrel+21}. The relevance here is that it follows mathematically from the very definition of the star formation rate. We now notice the following points:}

\begin{enumerate}

\item Eq.~(\ref{eq:fundamental}) is general. Thus, it is valid for any gas mass $\Mgas$ (molecular, atomic, or even ionized\footnote{Although ionized gas is not forming stars, we explicitly wrote `ionized' to stress that the equation is still valid mathematically even in such a case. Indeed, in this case, the `efficiency per free-fall time' will be zero, and so will the star formation rate, making the equation.}), independently of its physical state (collapsing, rotating, turbulent, magnetized, etc.). However, the existence of a correlation between the star formation rate and the ratio of gas mass over time will depend on the behaviour of $\epsilon$ and $\tau$, as we will see below.  

\item In deriving eq.~(\ref{eq:fundamental}), we have multiplied and divided by an arbitrary timescale. This could be the free-fall time, $\tauff$, the turbulent crossing time $\taudyn=l/\deltavl$, where $l$ is the scale length with a velocity dispersion $\deltavl$, the orbital time of a galaxy, $\tauorb=2\pi R/V_{\rm circ}$, or any other timescale. Nonetheless, the mathematical validity of the equation does not imply that there will be a linear correlation between $\promedio{\SFR}_{\tausf}$ and the ratio $\Mgas/\tau$.

\item Two different time scales are involved in eq.~(\ref{eq:fundamental}), which are not necessarily equal. The timescale $\tausf$ over which the star formation rate is averaged out, and the characteristic timescale of the fluid, $\tau$.

\item If $\epsilon$ were approximately constant, dividing both sides of eq.~(\ref{eq:fundamental}) by the volume and since the free-fall timescale is also related to the volume density, one can reproduce the Schmidt conjecture, i.e., { a correlation between (a measurement of) the star formation rate and some power-law of the volume density of gas mass. A similar relation can be found for the column densities if we divide by the area of the cloud.} 

\item If a linear function between (a measurement of) $\SFR$ and $\Mgas/\tau$ can be fitted to the data, then the ratio $\epsilon$ between the involved timescale $\tau$ and the depletion time $\taudepl$ has to be constant for such data.

\item If the characteristic timescale $\tau$ used in the derivation of eq.~(\ref{eq:fundamental}) is the free-fall time { of a mass that is actually collapsing, the factor $\epsilon$ becomes the so-called `efficiency per free-fall time' parameter,} $\eff$.

{
\item It is important to note that the `efficiency per free-fall time' parameter, in reality, is not an actual efficiency. We will discuss this point in \S\ref{sec:eff} and, for now, keep this name because it is quite rooted in the literature. However, we will write it in quotes from now on. }

\end{enumerate}

We thus rewrite eq.~(\ref{eq:fundamental}) in terms of the free-fall time { and the mass in gas that is collapsing, $\Mcollapsing$}
\begin{equation}
  \promedio{\mathrm{SFR}}_\tausf = \eff\ \bigg(\frac{\Mcollapsing}{\tauff}\bigg),
  \label{eq:fundamental2}
\end{equation}
 with
\begin{equation}
    \eff \equiv \frac{\tauff}{\taudepl} =\frac{\SFR}{\CR}\,,
    \label{eq:eff:definition}
\end{equation}
where $\CR$ is the rate at which the gas es collapsing,
\begin{equation} 
 \CR = \frac{\Mcollapsing}{\tauff}
\end{equation}
}

Dividing equation (\ref{eq:fundamental2}) { by the area or volume of the region under analysis, the surface- or the volumetric-law of star formation is recovered. These are equivalent. And since they all have} been derived from the sole definition of the star formation rate and basic algebra with no additional assumption in its derivation, and since gravity is a crucial ingredient for star formation, we will call it the fundamental law for star formation, { in agreement with previous works \citep[e.g., ][]{Krumholz+12, Elmegreen18, Pokhrel+21}.

A key feature of eq.~(\ref{eq:fundamental2}) is that, in observations, one is not completely sure what is the mass that is actually collapsing. As a first approximation, observational data from a single tracer will trace gas at a characteristic density and, thus, a characteristic free-fall time. This will provide, thus, a linear $\Sigmasfr-\Sigmagas$ star formation law, as is the case of CO or HCN observations. The inclusion of gas at different densities for deriving the star formation law, as is the case of {\it Herschel} observations of individual clouds \citep{Pokhrel+21}, or HI and molecular gas \citep[e.g., ][]{Kennicutt98, Bigiel+08, Kennicutt_DeLosReyes21, Sun+23, Ellison+24}, will provide correlations that are necessarily non-linear since the free-fall time is different at different densities. We discuss these issues in \S\ref{sec:galactic} and \ref{sec:extragalactic}.

}

\subsection{Is collapse the reason for the constancy of $\eff$?}\label{sec:eff_cst2}

{ Although $\eff$ varies over two orders of magnitude, from $\eff\sim3\times$\diezalamenos3 to 0.3 \citep[see Fig.~10 in][and references therein]{Krumholz+19}, it is frequently assumed that it is nearly constant, around $\eff\sim0.01$--0.03 \citep{Krumholz_McKee05, Krumholz+12, Elmegreen18}. In particular, \citet{Pokhrel+21} found $\eff\sim 0.026$ with a spread of 0.2~dex for a set of different clouds in the Solar Neighbrohood. In principle, one can show that if MCs are collapsing and thus undergoing Jeans fragmentation, then the expected values for $\eff$ do not vary too much for a large sample of clouds. For this purpose, we first notice that the total mass in newborn stars can be estimated as the product of the total number of stars multiplied by their mean mass,}
\begin{equation}
  \Mtotstarstausf = \Ntotstarstausf \promedio{M_*} .
\label{eq:MassNyso}
\end{equation}
Thus, the `efficiency per free-fall time'  (\ref{eq:eff:definition}) can be rewritten in terms of two constants: the mean mass of the newborn star, $\promedio{M_*}$, and the timescale in which the stars are formed. The justification for the first one is that the initial mass function is nearly universal and has a characteristic mass $\promedio{M}\sim 0.45$~\Msun\  \citep{kirkpatrick+24}. The justification for the second one is that the different evolutionary stages of a protostar have their characteristic timescales. For instance, if the objects to count are YSOs in general, the assumed characteristic timescale is $\sim$2~Myr. In contrast, when considering embedded protostellar objects, the characteristic timescale is of the order of 0.5~Myr \citep[e.g.,][]{Evans+09, Heiderman+10, Lada+10, Lada+13, Pokhrel+21}.

Having $\promedio{M_*}$ and $\tausf$ constant, we notice that the requirement for keeping $\eff$ constant, is that the factor 
\begin{equation}
  \frac{N_{*,\tausf}}{\Mgas} \tauff
  \label{eq:ratio}
\end{equation}
remains constant. Let us assume now that we have a super-Jeans, collapsing, nearly isothermal molecular cloud. After a given time, this cloud will form some number $N_*$ of stars. Suppose now that we vary the system's mass by an arbitrary factor $f$, keeping the same density (and thus, the same free-fall time). In that case, there will also be a factor $f$ of difference in the number of Jeans masses and, thus, one can expect, statistically speaking\footnote{By ``statistically speaking" we mean that different realizations of the same experiment may have different outcomes, but the number of protostars oscillates around a mean characteristic value.}, that the number of protostars will also vary by the same factor $f$, such that the ratio (\ref{eq:ratio}) remains constant. Mathematically, if the gas is super-Jeans and collapsing, one can expect the number of stars formed in a gravitationally-dominated { cloud to be proportional to the { inverse} of the Jeans mass \citep[see, e.g., ][]{Bate_Bonnell05, Palau+15, Morii+24},}
\begin{equation}
  N_* \propto \Njeans = \frac{\Mgas}{\Mjeans} ,
  \label{eq:Njeans}
\end{equation}
with
\begin{equation}
  \Mjeans \sim 2~M_\odot \ \bigg(\frac{c_s}{0.2~\mathrm{km\,s^{-1}}}\bigg)^3\ \bigg(\frac{n}{10^3~\mathrm{cm^{-3}}}\bigg)^{-1/2} .
\end{equation}
Since also $\tauff\propto n^{-1/2}$, then the ratio  (\ref{eq:ratio}) should not depend on the density, confirming the constancy of $\eff$, { provided the velocity dispersion $c_s$ is constant}.

We can make the reverse thought experiment. For this purpose, let us now take the same cloud, adding a factor $f$ in mass to maintain the same density. But now let us assume that the added gas is strongly stirred by supersonic turbulence with some virial parameter $\alphavir > 2$. In that case, the number of stars that will be formed in the same free-fall time will depend on ${{\alphavir}}$, as shown by, e.g., \citet[][]{Padoan+12} and \citet{Kim+21}. Consequently, the `efficiency per free-fall time' $\eff$ would not remain constant.

{ 

In what follows, a basic point to understand the different correlations is that the dense and cold gas collapses faster than the less dense and/or warmer gas. Thus, although the gas does not need to be molecular to collapse \citep{Clark_Glover14}, molecular gas tends to be colder and denser and then to collapse faster. Consequently, since clouds are density segregated \citep[e.g., ][]{Alfaro_Roman-Zuniga18, Morii+23}, in the presence of both cold, dense molecular gas and warmer and less dense atomic gas, the former will collapse faster and dominate the star formation. In a sense, molecular gas is a proxy for cold and dense gas prone to collapse, and consequently, it has to have a better correlation with the star formation than the warmer gas. 

We notice that considering tracers with different characteristic densities and free-fall timescales hides the fundamental law of star formation (\ref{eq:fundamental2}). Indeed, in a several-tracer measurement, the gas that mainly drives the correlation is the dense and cold, typically the molecular gas. Thus, including additional gas that will not end up forming stars will only hide the star formation law. In fact, it is worth noticing how the clear linear correlations between the $\SFR$ and the mass in molecular gas become messy when adding atomic gas \citep[e.g., ][]{Bigiel+08}.

}

\section{The Galactic K-S correlations as a consequence of the fundamental law}\label{sec:galactic}

The { demonstration} shown in \S\ref{sec:eff_cst2} allows us to understand the variety of correlations observed for the Milky Way and external galaxies quoted in\S\ref{sec:intro} with a single explanation, { as we discuss in what follows.}  

\subsection{The intra-cloud $\Sigmasfr-\Sigmagas/\tauff$ correlation}
\label{sec:SigmaSFR-Sigmagas_over_tau}

The correlation
\begin{equation}
 \Sigmasfr  = \eff\ \bigg(\frac{\Sigmagas}{\tauff}\bigg)
\label{eq:Pokhrel_correlation}
\end{equation}
observed for individual clouds by \citet{Pokhrel+21}, follows directly from the fundamental equation of star formation since it is obtained from eq.~(\ref{eq:fundamental2}) by simply dividing it by the projected area of the cloud. In this case, the constant of proportionality is the `efficiency per free-fall time'. 

\subsection{The Lada's inter-cloud correlation $\Nysomath-\Mdense$ }
\label{sec:Nyso-Mdense}

The correlation, 
\begin{equation}
  {  \Nysomath \propto \Mdense }
  \label{eq:lada}
\end{equation}  
shown by \citet{Lada+10}, is also a direct consequence of the fundamental law of star formation. The demonstration is simple. We first notice that in Solar Neighborhood’s clouds studies \citep[e.g., ][]{Evans+09, Heiderman+10, Lada+10, Gutermuth+11, Lombardi+13, Lombardi+14, Lada+17, Pokhrel+21}, observers typically estimate the mass in young stellar objects as { described in eq.~\ref{eq:MassNyso},} i.e., as the number of observed young stellar objects $\Nysomath$  multiplied by their mean mass, $\promedio{M_*}$ \citep[see also][] {Kennicutt_Evans12}. { In this case, the fundamental law of star formation can be written as:
\begin{equation}
  \promedio{\mathrm{SFR}}_\tausf = \frac{\Nysomath\,\promedio{M_*}}{\tausf} =
    \eff\ \bigg(\frac{\Mcollapsing}{\tauff}\bigg),
\end{equation}
with $\promedio{M_*}$ and $\tausf$ being constant for clouds under scrutiny.}
Furthermore, according to \citet{Lada+10}, the clouds in that study have nearly the same volume density. If so, these clouds should have nearly the same free-fall timescale $\tauff$. { Therefore, in the equation
\begin{equation}
 \Nysomath =
    \bigg(\frac{\eff\,\tausf}{\promedio{M_*}\,\tauff}\bigg)\, \Mcollapsing,
\end{equation}
the quantities inside the parenthesis are constant as long as $\eff$ remains constant.
}
Thus, if the clouds under analysis are collapsing, then the correlation $\Nysomath\propto\Mgas$ follows from the constancy of the `efficiency per free-fall time'. In a sense, this correlation is an example of our very deduction of the constancy of $\eff$, since different collapsing clouds may exhibit different amounts of Jeans masses.

\subsection{The Galactic intra-cloud $\Sigmasfr-\Sigmagas^2$ correlation for most local clouds, but $\Sigmasfr-\Sigmagas^{3.3}$ for California MC}\label{sec:SigmaSFR-Sigmagas2:new}

The previous correlations are { nothing but} the direct application of eq.~(\ref{eq:fundamental2}). We now want to understand the fact that, internally, most molecular clouds in the Solar Neighborhood do exhibit a power-law correlation between their surface densities of star formation and gas mass  (eq.~\ref{eq:KS-Pohkrel}), where most of the clouds show a power-law index of $N=2$ \citep[][]{Pokhrel+21, Spilker+21}, but the California MC exhibits $N=3.3$. The understanding of these correlations is more subtle since we want to make them work simultaneously along with eq.~(\ref{eq:Pokhrel_correlation}), and thus, it involves understanding how clouds' physical properties are measured and how they are related to the { internal structure of the cloud through its column density. Indeed, as \citet{Lada+13} showed, ``the cloud’s structure plays a fundamental role in setting the level of its star formation activity.''

With this idea in mind, we now want to find a relation between the power-law index of the column density PDF and the power-law index of the KS relation.} Let us assume then that $\eff$ is constant. We first notice that since 
\begin{equation}
  \tauff \propto \rho^{-1/2} \propto \bigg(\frac{M}{L^3}\bigg)^{-1/2},
\end{equation}
the KS relation (\ref{eq:Pokhrel_correlation}) can be rewritten in terms of the mass $M$ and size $L$ of the cloud, quantities that are directly read from the maps at different levels of the hierarchy:
\begin{equation}
  \Sigmasfr \propto \eff \frac{M^{3/2}}{L^{7/2}}.
\end{equation}
We further notice that if the column density of star formation rate is found to be proportional to some power-law of the  column density of the gas, $\Sigmasfr\propto \Sigmagas^N$, then 
\begin{equation}
  \frac{M^{3/2}}{L^{7/2}} \propto \frac{M^N}{L^{2N}}
  \label{eq:mass-size-constrain}
\end{equation}
Eq.~(\ref{eq:mass-size-constrain}) imposes a constraint on the observed mass-size relation, 
\begin{equation}
  M \propto L^{\alphauno}
  \label{eq:Mass_size_ratio_KS}
\end{equation}
which is 
\begin{equation}
  \alphauno = \frac{7-4N}{3-2N} .
  \label{eq:alphauno}
\end{equation}

We can explore further by recalling that the area $S$ (and thus its size $L\propto S^{1/2}$) and the cloud's mass $M$ depend on the column density PDF of the cloud, $P(\Sigmagas)$ as
\begin{equation}
  S(\Sigmagascero) \propto \int_{\Sigmagascero}^\infty P(\Sigmagas) d\Sigmagas
  \label{eq:surface:pdf}
\end{equation}
and
\begin{equation}
  M(\Sigmagascero) \propto \int_{\Sigmagascero}^\infty \Sigmagas\ P(\Sigmagas) \ d\Sigmagas 
  \label{eq:mass:pdf},
\end{equation}
with $\Sigmagascero$ the column density contour above which we define the cloud \citep[see][]{Lombardi+10, Ballesteros-Paredes+12}. Assuming that the   column density PDF $P(\Sigmagas)$ of collapsing clouds is a power-law \citep{Kainulainen+09, Kritsuk+11, Ballesteros-Paredes+11b, Ballesteros-Paredes+12, Girichidis+14}, and labeling $-q$ the power-law index,
\begin{equation}
  P(\Sigmagas)\ d\Sigmagas = \bigg(\frac{\Sigmagas}{\Sigma_{\rm gas, *}}\bigg)^{-q} d \Sigmagas
\end{equation}
where $\Sigma_{\rm gas,*}$ is only a normalization constant, one can prove that 
\begin{equation}
  {M}\propto L^\alphados ,
  \label{eq:Mass_size_ratio_PDF}
\end{equation}
with 
\begin{equation}
  \alphados= \frac{4-2q}{1-q} .
  \label{eq:alphados}
\end{equation}
Thus, eqs.~(\ref{eq:alphauno}) and (\ref{eq:alphados}) give us the relation between the power-law index of the KS relation, $N$, and the power-law index of the column density PDF, $q$, of a given cloud:
\begin{equation}
  q = 4N-5 .
  \label{eq:KS-PDF:relation}
\end{equation}

We can now get back to the KS correlations found in local clouds. On the one hand, with the possible exception of Cygnus-X and AFGL-490, each cloud in the \citet{Pokhrel+21} sample exhibits a quite straight $\Sigmasfr\propto \Sigmagas^2$ correlation, as well as a quite constant `efficiency per free-fall time' (see their Figs.~2 and 3) as a function of the column density. Given the fact that these clouds follow the KS-relation with $N=2$, eq.~(\ref{eq:KS-PDF:relation}) tells us that the power-law index of the PDF of these clouds must be $q=-3$. This is precisely the index found for the column density PDF of some of them, as shown by \citet{Lombardi+14} for Orion A and B clouds, and by \citet{Zari+16} for the Perseus molecular cloud. We speculate, thus, that the PDF of the column density for the remaining clouds (with the possible exception of Cygnus-X and AFGL-490, see above) observed with Herschel in the \citet{Pokhrel+21} study must follow nearly the same trend.

Let us now analyze the California cloud. Interestingly, this cloud is not present in the analysis of \citet{Pokhrel+21}, even though it is one of the closest and better-known clouds, and it has been observed by {\it Herschel}. Fortunately, both the column density PDF and the KS relation of this cloud were studied by these authors \citep{Lada+17}. They found a KS-type relation with $N=3.3$, i.e.,  $\Sigmasfr\propto\Sigmagas^{3.3}$. This implies that if the `efficiency per free-fall time' were constant, the power-law index of its PDF should be $q=8.2$, substantially steeper than the actual value of $q=3$ reported by \citet{Lada+17}. This implies that the California molecular cloud must not have a constant `efficiency per free-fall time' as a function of column density but an efficiency that should increase with column density. 

We speculate that the California cloud must have a non-constant `efficiency per free-fall time', and that this is consistent with the cloud being a more diffuse, more extended, and less dense cloud than, e.g., Orion A cloud, as found by \citet{Rezaei_Kainulainen22}. Indeed, the cloud has a mass comparable to Orion's A or B clouds \citep[e.g.,][]{Lada+10}. Still, Gaia analyses of the distances to the stars associated with California and Orion complexes suggest that the California cloud is substantially more extended in the line of sight and, thus, substantially less dense. {Furthermore, its star formation activity is substantially smaller than in other clouds of the \citet{Pokhrel+21} sample, hosting only one significant stellar group, LkH$\alpha$-101, in addition to a few sparse YSOs \citep{Lada+17}.} Thus, one can expect that not all the gas in the California cloud is as gravitationally bound and, thus, not necessarily involved in the process of collapse. If this is the case, there is no reason for the `efficiency per free-fall time' to be constant with column density.

\subsection{The lack of a correlation between $\Sigmasfr$ and $\Sigmagas$ for local clouds observed at a given column density threshold}

{ \citet{Lada+13} have shown that there is no correlation between the column densities of star formation rate and molecular gas mass for a set of clouds in the Solar Neighbrohood. These results appear to be at odds with their own correlation between the $\Nysomath\propto \SFR$ and $\Mgas$ discussed above, as well as with the} existence of a $\Sigmasfr \propto\Sigmagas^2$ correlation for the inner structure of molecular clouds studied by \citet{Pokhrel+21}. 
On the contrary, { at the same column density,} different clouds have a $\Sigmasfr$ that varies about one order of magnitude among them. The reason  
{ for this has been discussed by \citet{Krumholz+12}. At a given column density threshold, MCs may very well have different masses and densities. Thus, the star formation rate should have a natural scatter at every column density. In other words, the lack of a correlation found by \citet[][see their fig.~8]{Lada+13} is nothing but a single cut in column density of the result found by \citep[][see their Fig.~1]{Pokhrel+21}. The different`efficiencies per free-fall time' for each cloud result from a natural scatter in the mass and free-fall timescale involved in eq.~(\ref{eq:fundamental2}). Individual differences in the geometry and velocity field may also contribute to this scatter \citep[e.g., ][]{Hartmann_Burkert07, Zavala-Molina+24}.

\section{The extra-galactic K-S correlations as a consequence of the fundamental law}\label{sec:extragalactic}

Our original motivation was mainly { to give a simple, basic explanation for the variety of galactic correlations found in the literature \citep[e.g., ][]{Lada+10, Lada+13, Lada+17, Lombardi+14, Zari+16, Pokhrel+21}. However, being so basic, the fundamental law of star formation (\ref{eq:fundamental2})  can also explain the variety of extragalactic correlations, as we discuss in what follows.  
Similar and/or alternative explanations have already been provided by \citet{Elmegreen18}. We discuss this issue in \S\ref{sec:whatsnew}.
}

\subsection{The Gao-Solomon correlation}

 The fundamental equation of star formation (\ref{eq:fundamental2}) shows a natural and immediate explanation for the \citet{Gao-Solomon04} correlation and its extension shown by \citet{Wu+05}. Since for that correlation, massive dense cores were observed using the same tracer, HCN, we can assume that all of these cores have nearly the same density at which the HCN is excited, regardless of its actual value\footnote{Note, however, that recent suggestions that HCN may be sub-thermally excited in especially active star-forming regions \citep[e.g., the LEGO Survey][]{Barnes+20}.}. Thus, we can expect $\tauff$ to be nearly the same for all clouds. Finally, since HCN is a tracer of dense gas $(n\sim 10^4$~cm\alamenos3), we can assume that most of that gas is in the process of collapse, producing thus the correlation. This is similar to the argument we gave in \S\ref{sec:Nyso-Mdense} to explain the correlation found by \citet{Lada+10} between the number of YSOs and the mass in dense gas.

\subsection{The extragalactic $\Sigmasfr-\SigmaMC$ correlation { (the mKS correlation)}}\label{sec:extragalactic:Ssfr-Smc}

An important ingredient in understanding this correlation is that even though { MCs have a wide range of densities distributed in a 
hierarchical way} (i.e., they have internal structure), the filling factor of the dense structures is so small\footnote{Another way of picturing this fact is that the column density PDFs of molecular clouds decrease fast.} that the mean density of a cloud is always at the lower levels of its hierarchy, a prediction made more than ten years ago \citep{Ballesteros-Paredes+12} and confirmed observationally recently by  \citet{Cahlon+23}. Thus, since this correlation has been performed using CO \citep[e.g., ][]{Bigiel+08, Sun+23, Jimenez-Donaire+23, Ellison+24}, and since CO is excited at around $100-400$~cm\alamenos3, the observed CO clouds have mean densities around these values. Consequently, { all CO clouds have} similar free-fall timescales, and thus, the linear correlation between the column densities of star formation and of molecular gas follows naturally from the fundamental law of star formation. In other words, this correlation is explained in the same terms we explained the \citet{Gao-Solomon04}, \citet{Wu+05} and \citet{Lada+10} correlations: we are analyzing collapsing gas that traces basically the same density and thus has the same free-fall time.} 

{ 
\subsection{The shift in $\Sigmasfr$ between CO and HCN correlations}

A reassuring feature regarding the relevance of eq.~(\ref{eq:fundamental2}) is the displacement of one order of magnitude in $\Sigmasfr$ between the correlations obtained using CO with typical densities of 100~cm\alamenos3 \citep[e.g., ][]{Bigiel+08, Sun+23, Ellison+24}, and those using gas that is at 10\ala4~cm\alamenos3 \citep{Gao-Solomon04, Wu+05, Lada+10}. This displacement of a factor of 10 in $\Sigmasfr$ is consistent with CO collapsing in timescales 10 times longer than the HCN clouds and the \citet{Lada+10} clouds, which have densities 100 times larger.

}

\subsection{The super-linear correlation $\Sigmasfr-\Sigmagas$}
\label{sec:superlinear}

The explanation of the { super}-linear KS relation for resolved galaxies \citep{Bigiel+08, Leroy+08}, when using total gas column density (HI $+$ H$_2$) { has been already hinted in \S\ref{sec:eff_cst2}. The fundamental point is that atomic gas in the ISM is typically less dense and warmer than molecular gas, and thus, it is less prone to collapse. In addition, { once self-shielding against the dissociating UV radiation is achieved,} the mass fraction of molecular gas correlates with total surface density \citep[e.g., ][]{Park+23}, because the larger the column density, the larger the self-shielding against the UV dissociating radiation \citep{Hartmann+01, Bergin+04}.

Let us assume, thus,  that we can first measure a particular star formation rate for a small column density of total} gas, $\Sigma_{\rm gas,1}$, and that at these column densities, the molecular gas fraction $f_{\rm mol, 1}$ is smaller than that of the HI, $f_{\rm HI, 1}$. We now observe another parcel of the galaxy where the column density is larger by, e.g.,  a factor of two, $\Sigma_{\rm gas,2} = 2\ \Sigma_{\rm gas,1}$. { We know that the larger the density, the more easily the cloud is shielded against the dissociating UV radiation, and the more likely the H$_2$ can be formed.} Consequently, not only the mass of H$_2$ does increase, but it does it at the expense of the HI, i.e., the fraction of molecular gas also increases such that $f_{\rm mol,1} < f_{\rm mol,2}$. In other words, although we have doubled the total column density, the amount of molecular gas in the second case is larger than twice its original value. Thus, even if the column density of total gas (HI+H$_2$) is twice the former, the column density of star formation increases faster because the column density of { cold and dense} molecular gas has also increased faster, producing the super-linear power-law for the K-S correlation.

{ It should be stressed, nonetheless, that although the global correlation between the column densities of star formation and total gas (HI+H$_2$) is superlinear, in the regime $\Sigmagas>$~10~$M_\odot$~pc\alamenos2, i.e., in the regime where typically the gas becomes predominantly molecular, \citep[see, e.g.,][]{Hartmann+01, Park+23}, the tendencies shown by \citet[][ see the right panels of Fig.~4]{Bigiel+08} become flatter, substantially close to unity.

}

\subsection{The lack of a correlation at { sub-kpc} scales}

As commented in the introduction, the correlation is broken, or at least more sparse, for extragalactic resolved objects at smaller scales $\lesssim$0.5~kpc as has been observed by, e.g., \citet[][]{Schruba+10, Williams+18, Pessa+22, Pan+22}. As these authors explained, some of their target fields may focus on a molecular cloud that obscures the H$_\alpha$ emission, accounting for smaller $\Sigmasfr$ and larger $\Sigmagas$. At the same time, other beams may point to regions with more significant H$_\alpha$ emission but few or no CO emission \citep{Schruba+10}. This means that, given the methodology in extragalactic studies, small volumes have sampling problems. Extragalactic observations require large averaging volumes to track simultaneously the star formation and the presence of molecular gas.

\subsection{The sub-linear $\Sigmasfr-\SigmaMC/\tauff$ correlation for galaxies}\label{sec:SFR_Sigma_over_tauff}

The result provided by \citet{Sun+23},
\begin{equation}
  \log{\Sigmasfr} = \alpha + \beta\log{\SigmaMC/\tauff}
\end{equation}
with $\beta\sim 0.7$, appears to contradict our analysis that this correlation should be linear because molecular gas is in the state of collapse, { and CO clouds have basically the same volumetric density}. This inconsistency has its origin, however, in the way in which \citet{Sun+23} estimated column densities and free-fall times of molecular gas. These authors computed the column density at 1.5~kpc resolution by integrating the CO(2-1) intensity from the moment-0 maps,  and assume a common line-of-sight length of $l=$150~pc for computing the { volumetric} density,
\begin{equation}
  \rho \propto {\Sigma}/{l}
\end{equation}
and thus, the free-fall time \citep[see eqs.~(3) and (4) in][]{Leroy+16}. This methodology introduces the bias that produces the sub-linear power-law. To understand this point, we first recall that molecular clouds are highly fractal \citep{Falgarone+91} and have sizes in the range $0.1-100$~pc \citep[e.g., ][]{Larson81, Solomon+87, Blitz+07}. This means that the filling factor of MCs within 1.5~kpc size boxes is substantially small. Imagine that we have a (1.5~kpc)$^2$ box of a disk galaxy containing only one CO cloud, with a size of, say, $l\sim$~60~pc. Its mean density will be around a few hundred cm\alamenos3, and its free-fall time will be around three million years. Let us assume now that we have ten other clouds in the same (1.5~kpc)$^2$ box. The total mass in the box will be ten times larger, and so it will be the estimated column { and volumetric densities. However, the real mean density of molecular gas will still be a few hundred cm\alamenos3, because all the clouds considered have the same density. As it can be seen, the methodology used by \citet{Sun+23} introduces a bias: for each order of magnitude larger column density, the estimated free-fall time will be a factor $\sqrt{10}\sim 3$ shorter. } Consequently, the one-to-one correlation between $\Sigmasfr$ and $\Sigmagas$ implies that the power-law index of the $\Sigmasfr-\Sigmagas/\tauff$ correlation will be 1/3 smaller because we are { stretching the $x$ axis by a factor of $\sim${3}}. Indeed, the power-law found by \citet{Sun+23} exhibits power-law indexes of about 2/3 \citep[see the third section of Table~2 of][]{Sun+23}, { confirming our explanation.}

\subsection{The sub-linear { Elmegreen-Silk} correlation $\Sigmasfr-\SigmaMC/\tauorb$} \label{sec:Elmegree-Silk}

\begin{figure}
  \includegraphics[width=0.95 \columnwidth]{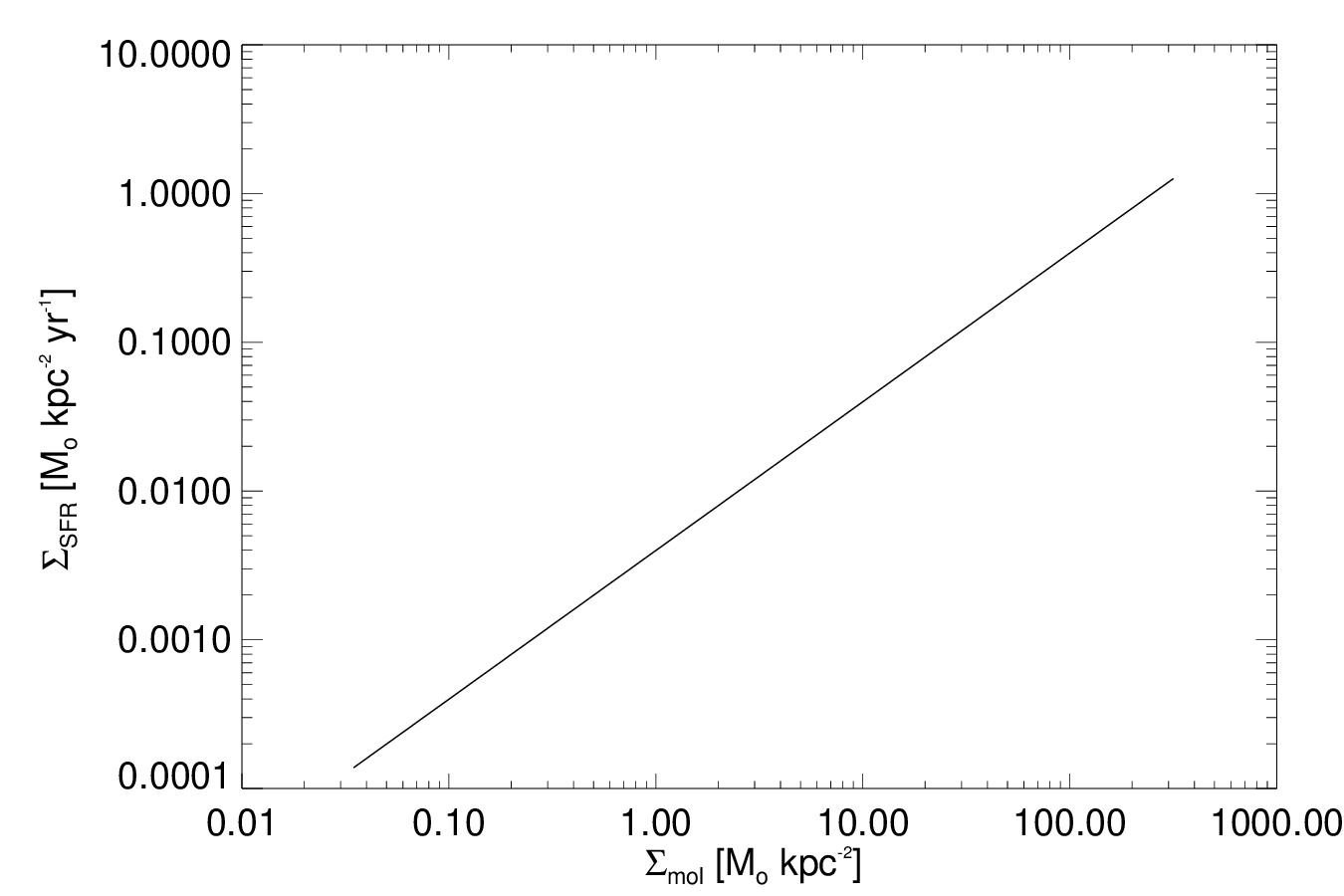}
  \includegraphics[width=\columnwidth]{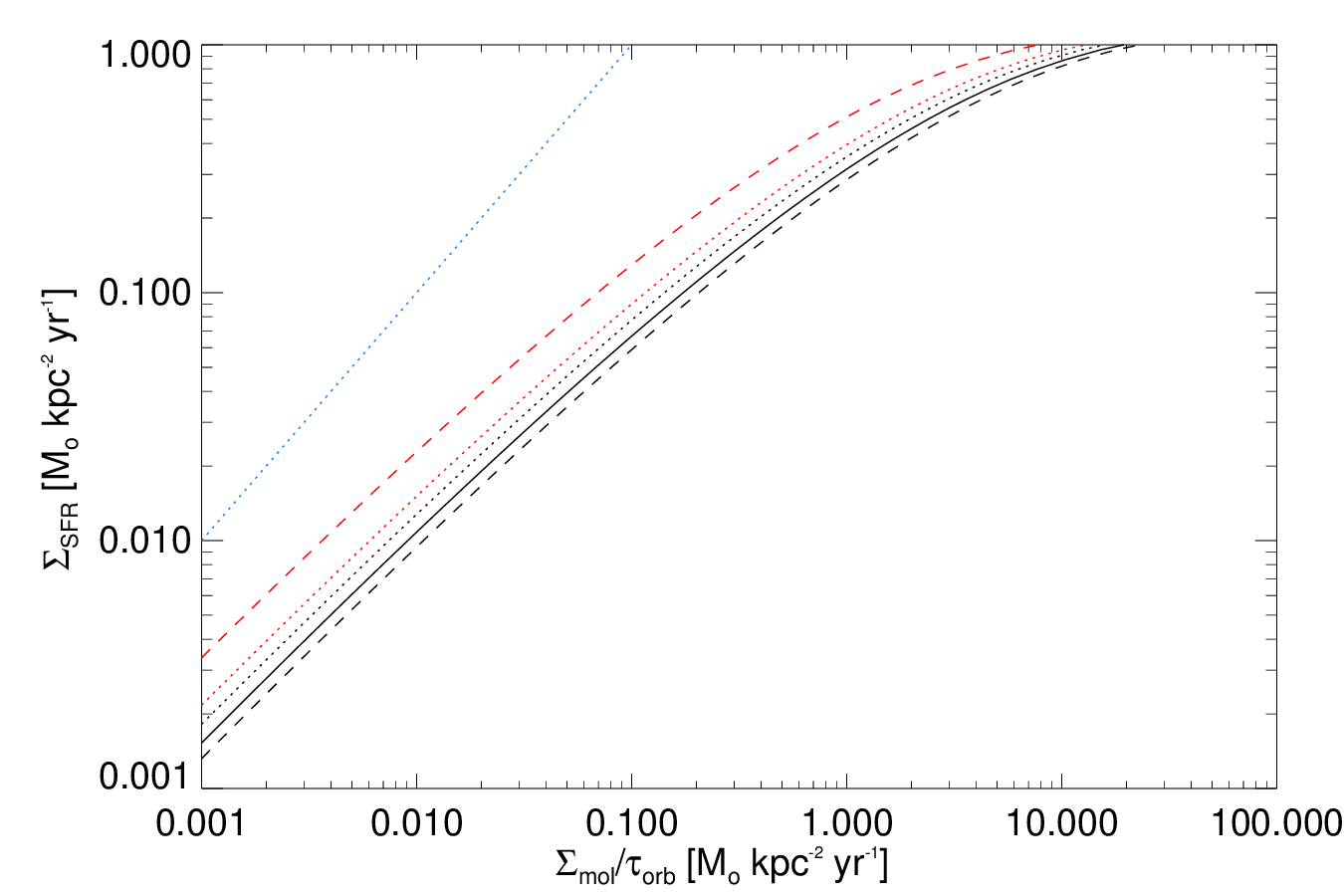} \\
  \caption{{Kennicutt-Schmidt diagrams. In the upper panel, we show the power-law fit found by \citet{Sun+23} for a sample of 80 galaxies. In the lower panel, we have assumed that the column density of the molecular gas in galaxies has a decreasing exponential form and that the galaxies have flat rotation in order to compute the $\Sigmasfr-\SigmaMC/\tauorb$. These figures can be compared to Figs.~ 1 and 2 of \citet{Sun+23}. As it can be seen, the correlation becomes sublinear when dividing by $\tauorb$, and the power law index is around 2/3, consistent with \citet{Sun+23}.}}
  \label{fig:E-S}
\end{figure}

{

As mentioned before, \citet{Sun+23} found a sublinear Elmegreen-Silk correlation \citep{Elmegreen97, Silk97},
\begin{equation}
   \Sigmasfr \propto \bigg(\frac{\SigmaMC}{\tauorb}\bigg)^N .
   \label{eq:KS-orbital}
\end{equation}
with $N\sim 0.8$. These authors already explained the origin of this correlation, arguing that the orbital time $\tauorb$ is smaller in the inner regions of galaxies, where the column density is larger, stretching the correlation to the right. To quantify the level of stretching, we first} recall that the orbital time used in the correlation is related to the circular velocity $V_{\rm circ}$ as
\begin{equation}
  \tauorb = \frac{2\pi R}{V_{\rm circ}} ,
  \label{eq:tauorb}
\end{equation}
and that the column density of gas in galaxies typically has a decreasing exponential shape 
\begin{equation}
  \SigmaMC = \Sigma_0 \exp\bigg(-{\frac{R}{R_{\rm 0, mol}}}\bigg)
  \label{eq:Sigmagas}
\end{equation}
where $R_{\rm 0,mol}$ is the scale length of the column density of the gas, { and $\Sigma_0$ is the central density of the exponential \citep[for the CO clouds in the Milky Way,][estimate $R_{\rm 0, mol}\sim2$~kpc, and $\Sigma_0\sim100$~\Msun~pc\alamenos2]{Miville-Deschenes+17}.} Solving for $R$ in (\ref{eq:Sigmagas}) and substituting it in (\ref{eq:tauorb}), one obtains
\begin{equation}
  \tauorb \ = \ -~2\pi~\bigg(\frac{R_{\rm 0, mol}}{V_{\rm circ}} \bigg)\  \ln\bigg({\frac{\SigmaMC}{\Sigma_0}}\bigg).
  \label{eq:tauorb2}
\end{equation}
{
Close to the centre of the galaxy, $\SigmaMC\to\Sigma_0$. Far from the center, i.e., $R\in\ \sim~(4,20)$~kpc, the ratio ${\SigmaMC}/{\Sigma_0}$ decreases from \diezalamenos1 to \diezalamenos3. Although the slope of the logarithmic function changes continuously, in this column density regime, we notice that
\begin{equation}
  -\ln{\bigg(\frac{\SigmaMC}{\Sigma_0}}\bigg) \simeq  1.3 \bigg(\frac{\SigmaMC}{\Sigma_0}\bigg)^{-0.25}.
  \label{eq:tauorb_approx}
\end{equation}
This means that by dividing $\SigmaMC$ by $\tauorb$, in practice we will compute $\SigmaMC^{1.25}$ instead of $\SigmaMC$ for each $\Sigmasfr$. Thus, we are stretching the $x$ axis from $\SigmaMC$ to $\SigmaMC^{1.25}$, which will convert the linear correlation into a sublinear power law, with a power-law index of the order of $\sim1/1.25=0.8$,
\begin{equation}
  \Sigmasfr\propto \SigmaMC^{0.8}
\end{equation}
which is approximately the reported correlation by \citet{Sun+23} for their sample of galaxies.

To illustrate the effect of dividing $\SigmaMC$ by $\tauorb$, we can then build a variable $\SigmaMC\in  (0.1,100)$~\Msun~pc\alamenos2, i.e., in the range in which the CO is measured,} and with it, compute $\Sigmasfr$ from
\begin{equation}
  \log{\Sigmasfr} = \alpha \ +\ \beta\ \SigmaMC
  \label{eq:fiducial}
\end{equation}
using the fiducial values of $\alpha\sim{-2.4}$ and $\beta\sim 1$ obtained by \citet{Sun+23} {  for their mKS case.  Fig.~\ref{fig:E-S} shows, in the upper panel, $\Sigmasfr$ vs $\SigmaMC$ as given by eq.~(\ref{eq:fiducial}). In the lower panel we plot $\Sigmasfr$ vs $\SigmaMC/\tauorb$, using eq.~(\ref{eq:tauorb2}) for the $x$ axis.} The black solid line was computed using the canonical values of the Milky Way: $V_{\rm circ}= 220$~\kms, and $R_{\rm 0,mol}= 2$~kpc. For the black dotted and dashed lines, we used $V_{\rm circ}=$~180 and 260~\kms, respectively, keeping the same $R_{\rm 0,mol}$, while for the red dashed lines, we used $R_{\rm 0,mol} = $~3 and 5~kpc, keeping $V_{\rm circ}=220$~\kms. The blue dotted line is a power-law with an index of one. These figures must be compared to Figs.~1 and 2 of \citet{Sun+23}.

From Fig.~\ref{fig:E-S} it is clear { that the fiducial values for the Milky Way can reproduce the values reported by \citet{Sun+23},  and that the correlation becomes sublinear not because there was something intrinsic about the orbital time, but because it has a radial dependency, as well as $\SigmaMC$, and then, we stretch the $x$ axis when going from the molecular KS relation to the Elmegreen-Silk relation.} 

{ As can be seen,} this correlation can also be explained in terms of collapsing clouds. It should be stressed, furthermore, that although the correlation appears to be a power law, in reality, { it is not, although, in the range of column densities $\SigmaMC\in (1-100)$, it can be approached as a power law with index 0.8.} This implies that $\eorb$ is not constant, as pointed out by \citet{Sun+23}. { We also} notice that the inferred ``efficiency per orbital time'', or orbital-to-depletion timescale, is of the order of 0.05, i.e., since the mean depletion time is $\sim 2.5$~Gyr, the typical orbital time in those galaxies is of the order 125~Myr, although it varies from galaxy to galaxy and from place to place within the same galaxy.

{ 

\subsection{The slightly sublinear correlation between the star formation rate and the dynamical pressure $\Sigmasfr$-$\PDE$ correlation} \label{sec:pressure}

It has been argued that star formation may be regulated by the necessary pressure in the midplane of disk galaxies to achieve (vertical) dynamical equilibrium \citep[e.g, ][]{Blitz_Rosolowsky04, Blitz_Rosolowsky06, Ostriker+10}. The idea is quite interesting, { and it is schematically depicted in \citet[][Fig.~1]{Ostriker+10}. Here, we offer a summarized explanation:} if the ISM is overpressurized, it will compress the diffuse gas, favouring the formation of dense, cold clouds. This will lower the typical pressure of the ISM. By the same token, the larger amount of cold gas will increase the star formation rate, increasing the UV radiation (i.e., increasing the heating) and, thus, increasing the mean pressure of the thermal equilibrium curve. Then, the ISM de-pressurizes itself, and the thermal equilibrium curve increases its mean pressure. The reverse argument also applies: if the ISM is underpressurized, dense clouds will expand, increasing the mean pressure of the ISM. The star formation will decrease, reducing the UV radiation (i.e., reducing the heating) and then reducing the mean pressure of the thermal equilibrium curve. In this way, \citet{Ostriker+10} propose that star formation regulates itself. 

This pressure, which must correlate with the star formation rate, must be the pressure in the midplane due to the weight of the gas {. A commonly used approximation, which according to \citet{Ostriker_Kim22}, is accurate within 20\%} can be written as
\begin{equation}
 \PDE = \frac{\pi G}{2}   \Sigmagas^2 + \Sigmagas \sqrt{2G\rho_*} \sigmagasz ,
 \label{eq:presion}
\end{equation}
where {$\sigmagasz$ is the velocity dispersion of the gas perpendicular to the plane of the disk,} $\Sigmagas$ is the column density of the gas, and $\rho_*$ is the stellar density in the midplane \citep[{ note that we did not consider the contribution of the dark matter to the vertical structure of the disk, since we have found it to be negligible,}  e.g., ][]{Suarez+12, Ramirez-Galeano+22}.

Different authors have found, indeed, a correlation between $\Sigmasfr$ and $\PDE$ \citep[e.g., ][]{Barrera-Ballesteros+21, Sun+23, Ellison+24}. In particular, in their recent survey of~$\sim80$~galaxies, \citet[][]{Sun+23} have shown that the midplane pressure (\ref{eq:presion}) exhibits a slightly sublinear correlation with the column density of star formation rate, with an intrinsic scatter around the correlation not much different to the scatter given by the column densities of star formation rate and molecular gas mass.

There are several caveats to this approach. On the one hand, star formation is a process that occurs on scales of some Myr \citep[e.g., ][]{Leisawitz+89, Ballesteros-Paredes+99b, Hartmann+01, Ballesteros-Paredes_Hartmann07, Kruijssen+18, Chevance+20b}. In contrast, considering that the scale length of the HI can be of the order of $l_z\sim300$~pc \citep{Kalberla_Kerp09}, and the velocity dispersion of the order of $\sigmagasz\sim$7-11~\kms\ \citep{Heiles_Troland03, Haud_Kalberla07, Caldu-Primo_Schruba16}, the dynamical time $\taudyn = l_z/\sigmagasz$ is of the order of some tens of Myr, substantially larger than the free-fall time. { Then, although vertical equilibrium must be relevant for the assembly of dense clouds, it appears} irrelevant for regulating star formation itself.

In addition, at least for the Milky Way, the interstellar pressure is of the order of $P/k\sim3000$~K~cm\alamenos3, and thus, the required pressures for reaching dynamical equilibrium, $\PDE/k\sim$~10\ala4--10\ala6~K~cm\alamenos3 \citep[see, e.g., ][]{Barrera-Ballesteros+21, Sun+23, Ellison+24} are substantially larger. { While it is true that the total interstellar pressure (thermal, magnetic and cosmic ray pressure) in the Solar Neighborhood is ten times larger than the thermal pressure \citep[e.g.,][]{Boulares_Cox90}, \citet{Ostriker+10} recognize that the scale height of the magnetic field and cosmic rays are substantially larger than the scale height of the thermal pressure, making fields and cosmic rays negligible for vertical force balance. This suggests that, in order to achieve vertical equilibrium due to changes in the thermal pressure, average midplane thermal pressures have to increase by several orders of magnitude, which may occur locally (e.g., inside a supernova explosion), but will be hard to meet on kiloparsec scales.} 

Our interpretation of this correlation is that, with eq.~(\ref{eq:presion}), we are { mostly} sampling the same linear correlation $\Sigmasfr-\SigmaMC$ for molecular gas, disguised by additional factors. {
To understand this argument, we notice three points:}

\begin{enumerate}

{
\item As it can be seen, in Table~\ref{tab:pressures} we show values of the quadratic and linear terms (second and third columns) of eq.~(\ref{eq:presion}). From this Table one can notice that the quadratic term is negligible for densities below $\sim$~30~\Msun~pc\alamenos2, compared to the linear term, and it starts dominating at $\SigmaMC\gtrsim50$~\Msun~pc\alamenos2. However, since most of the data points in \citet{Sun+23} work are below 50~\Msun~pc\alamenos2, one can conclude that the quadratic term is typically negligible, but at the central regions of the galaxies, where the column densities reach large values.}

\item { It should be noticed, furthermore, that} although by construction, the total column density (HI+H$_2$) is used, the column density of H~I rarely contributes above $\Sigmagas > 10$~\Msun~pc\alamenos 2 \citep[see, e.g., ][]{Wong_Blitz02, Bigiel+08}. The reason is that above such column densities, most of the gas becomes molecular  \citep[e.g., ][]{Franco_Cox86, Hartmann+01, Bergin+04}. Thus, { although HI could play a role in $\PDE$, the actual values of $\SigmaMC$ and $\PDE$ in figs.~1 and 4 of \citet[][]{Sun+23} indicate that it does it only through the linear term, for column densities below $\Sigmagas\lesssim$10~\Msun~pc\alamenos2. Thus, the relevant column density in eq.~(\ref{eq:presion}) for most of the points is that of the molecular gas } i.e., $\Sigmagas\sim\SigmaMC$. In other words, eq.~(\ref{eq:presion}) is, in practice, probing that $ \PDE\propto\SigmaMC$ { for most of the data points}.

\item { As for the additional factors in eq.~(\ref{eq:presion}) entering in the linear term of} the gas column density (the velocity dispersion of the gas $\sigmagasz$ and the square root of the mid-plane stellar density) { we notice that, in practice, they do not vary substantially}. Indeed, as for the first factor, \citet{Caldu-Primo_Schruba16} have shown that it exhibits almost no variation as a function of radius along a handful of galaxies. In fact, \citet{Sun+23} uses a { fixed} value of $\sigmagasz=$11~\kms\ for all galaxies. As for the second factor, i.e., the square root of the stellar volume density, $\rho_*^{1/2}$, we noticed that for typical values of galaxies, this factor varies typically by less than one order of magnitude \citep[see, e.g., ][]{Barrera-Ballesteros+23}. Indeed, the stellar volume density is computed as $\rho_*\propto\Sigma_*/H_*$, with $H_*$ being a constant factor proportional to the radial scaling factor $R_*$. Since galaxy disks exhibit decreasing exponential column density profiles with scaling factors of the order of\footnote{ We found values of $R_*$ between 1--6~kpc for the sample in \citet[][see their Table~1]{Sun+20}.} $R_*\sim$2.5--4~kpc \citep[e.g., ][]{Drimmel_Spergel01, Chang+11, Mosenkov+21, Barrera-Ballesteros+23}, and since most of the CO and Ly$\alpha$ emission detected for exponential disks comes from radii between 3 and 12~kpc, in such a range, the square root of the stellar density varies by less than a factor of 10 \citep[see Fig.~3 in][]{Barrera-Ballesteros+23}. Thus, the second term of eq.~(\ref{eq:presion}) is basically proportional to the column density, with a slight stretching in the $x$ direction provided by the variation of $\Sigma_*^{1/2}$, which originates the slightly sublinear power-law in the $\Sigmasfr-P$ diagram.

\end{enumerate}

\begin{table}
\caption{Gas column density (first column) and approximate midplane dynamical equilibrium pressures due to the self-gravity of the gas (second column) and gravity of the stellar disk (third column).}
\label{tab:pressures}
\begin{center}
\begin{tabular}{ |c|c|c| } 
 \hline
$\Sigmagas$  & 
$\pi G \Sigmagas^2/2$     &
$(2 G \sigma_*)^{1/2} \sigma_{\mathrm{gas}, z} \ \Sigmagas$ \\
$M_\odot$~pc\alamenos2 & $k_B$ K cm\alamenos3 & $k_B$ K cm\alamenos3\\
 \hline
1       &33      & 1,570       \\
3       &296     & 4,710         \\
10      &3,288   & 15,700       \\
30      &29,595  & 47,097         \\
50      &82,209  & 78,495         \\
100     &328,836 & 129,727        \\
150     &739,881 & 194,592        \\
 \hline
\end{tabular}
\end{center}
\end{table}

{
We conclude that, by measuring $\PDE$ through the column density of gas with eq.~(\ref{eq:presion}), what we are doing is not estimating the actual total pressure in the midplane but estimating the midplane gravitational energy density due to the weight of the galaxy. In order to measure the actual midplane pressure, densities, temperatures, magnetic fields, and cosmic ray energy density should be measured. That will be the actual quantity to compare to $\Sigmasfr$. Otherwise, the $\Sigmasfr-\PDE$ correlation will only show how the $\Sigmasfr$ correlates with (a) $\Sigmagas$ at low column densities, (b) $\SigmaMC$ at middle column densities, and $f1\SigmaMC^2 + f2 \SigmaMC$ (with $f1$ and $f2$, factors that do not vary substantially with $\SigmaMC$) for very high column densities. Since most of the data points are at middle-column densities, it is not surprising that the slope of the $\Sigmasfr-\PDE$ correlation is close to linear.

In summary,} the $\Sigmasfr-\PDE$ correlation is { mostly} the fundamental equation of star formation, disguised by additional factors that have only a small variation with $\SigmaMC$.

\subsection{The \citet{Kennicutt98} laws for whole galaxies} \label{sec:kennicut98}

\citet{Kennicutt98} presented two correlations. The so-called Kennicutt-Schmidt relation 
\begin{equation}
  \Sigmasfr-\Sigmagas^{1.4},
  \label{eq:kennicut98}
\end{equation}
and the so-called Elmegreen-Silk relation,
\begin{equation}
  \Sigmasfr-\Sigmagas/\tauorb
  \label{eq:elmegreen-silk}
\end{equation}
\citep{Elmegreen97, Silk97}, both discussed above in terms of resolved galaxies at scales of $\sim1$~kpc, and using total (HI+H$_2$) gas. We interpret these two correlations in a similar way as before. As for the first one, the actual, linear correlation between the star formation rate and gas mass density is with the dense, cool molecular gas. However, the correlation becomes superlinear because of the inclusion of more diffuse, warmer atomic gas. This gas either does not collapse or does it in longer timescales because{, statistically speaking,} it is warmer and more diffuse. However, the larger the column density, the more easily the molecules are formed at expense of the atomic gas \citep[e.g., ][]{Bergin+04}, and then, as before (\S\ref{sec:superlinear}), a linear increase in the total column density of the galaxy translates it into a superlinear increase of the total cold, dense molecular gas, and consequently, in a superlinear increase of the column density of star formation rate.

As for the second correlation, we recall that
\begin{equation}
  \tauorb\propto\frac{R_{0,{\rm mol}}}{V_{\rm circ}} \ln\bigg(\frac{\SigmaMC}{\Sigma_0}\bigg) 
\end{equation}
(see eq.~(\ref{eq:tauorb2})). In this case, we just have to realize that we are analyzing global properties, and thus, each galaxy will have a single value for $\tauorb$, such that we will use again eq.~(\ref{eq:tauorb2}) for the ensemble of galaxies.

With this in mind, we first notice, as before,  that the ratio $\SigmaMC/\Sigma_0$ is quite constant for the variety of galaxies \citep[see e.g., ][]{Barrera-Ballesteros+23}. On the other hand, it is likely that the ratio $R_{0,{\rm mol}}/V_{\rm circ}$ remains constant for the ensemble of galaxies or does not vary more than a factor of two for most of them. To picture this, we notice that, in fact, the scale length of the stellar disk exhibits a linear correlation with the circular velocity \citep[][]{Courteau+07}, which implies a constant orbital time, close to 1 Gyr \citep{Meurer+18}, with a scatter dominated by the intrinsic scatter of the spin parameter of the configuration \citep{Mo+98, Hernandez_CervantesSodi06}
and thus, one can expect a similar result between the scale length of the molecular gas,  $R_{0,{\rm mol}}$ and the circular velocity $V_{\rm circ}$.

To further understand the Elmegreen-Silk correlation for whole galaxies as shown by \citet[][see his Fig.~7]{Kennicutt98}, we now note that, in reality, the regular galaxies (solid circles) exhibit a behaviour steeper compared to circumnuclear regions (open circles) and starburst galaxies (squares). Still, the scatter is such that all the points seem to follow the same global tendency. Consequently, we analyze regular galaxies independently from the starburst galaxies and circumnuclear regions. 

As it can be guessed, the orbital times of the starburst and circumnuclear regions are substantially smaller than those of regular galaxies. This means that in transiting from eq.~(\ref{eq:kennicut98}) to (\ref{eq:elmegreen-silk}),  we will stretch more the $x$ axis for the starburst galaxies and circumnuclear regions than for regular galaxies. In addition, the density contrast $\Sigmagas/\Sigma_0\sim 0.2-0.05$ for starburst and circumnuclear regions is substantially larger than that for disk galaxies. In this regime, $-\ln{\Sigmagas/\Sigma_0} \sim (\Sigmagas/\Sigma_0)^{-0.4}$. Thus, we will stretch the $x$ axis from $\Sigmagas$ to $\Sigmagas/\tauorb\propto \Sigmagas^{1.4}$ for the same range of $\Sigmasfr$. The new dependency becomes linear since the original dependency was already $\Sigmasfr\propto\Sigmagas^{1.4}$.

The above analysis implies that disk galaxies, for which we have found that $-\ln\Sigmagas/\Sigma_0\propto (\Sigmagas/\Sigma_0)^{-0.25}$, should exhibit a superlinear power-law index. Indeed, by looking only at the solid points in Fig.~7 of \citet{Kennicutt98}, one can clearly see that they do have an index larger than unity, although these authors did not notice it since the whole correlation is skewed by the tendency of the circumnuclear and starburst points, a fact noticed by \citet[][see their Fig.~3]{Kennicutt_DeLosReyes21}.
}

{
\section{Discussion}\label{sec:discussion}

\subsection{Gravity or turbulence? Collapse, overvirial CO clouds, and the KS relation}

In the previous sections, we have presented arguments to understand the different forms of the Kennicutt-Schmidt correlation in terms of a single star formation law that rises naturally as a consequence of the collapse of the gas under analysis. This is at odds with the common belief that turbulence supports clouds over many free-fall timescales and regulates the star formation.

Indeed, one of the main arguments against the idea of molecular clouds collapsing within a (few) free-fall times is that the time to exhaust all molecular gas in the galaxy is of the order of 1 Gyr, i.e, $\sim$300 times the free-fall time of a CO cloud. This { number has been used to argue} that star formation is a slow process \citep[e.g., ][]{Zuckerman_Evans74, Blitz_Shu80, Shu+87, Krumholz_Tan07, Kennicutt_Evans12, Krumholz+19, Evans+21, Evans+22}. In { such} scenario, molecular clouds last many free-fall times, forming stars at a low pace, and CO clouds are mostly supervirial \citep[e.g., ][]{Evans+21}, confined by external pressure \citep[e.g., ][]{Keto24} and, thus, not collapsing. 

To support this argument, it also has been argued that { even} those clouds exhibiting virial parameters around $\alphavir\sim$1--2, cannot be granted to be self-gravitating, since there are so many approximations in the estimations of the gravitational energy \citep{Chevance+23}. In addition, these authors made the argument that, while turbulence scales with size as $\sigma_v^2 \propto l$, where $\sigma_v$ is the velocity dispersion of the cloud at the characteristic size $l$, gravitational energy scales as $M/l\propto l^2$, implying that the virial parameter would scale as $\alphavir\propto l^{-1}$, and thus, smaller scales will be necessarily supported by turbulence. In other words, \citet{Chevance+20b} argue that clouds with virial parameters around unity are not necessarily bound, but even if they were, smaller portions of them would be supported against collapse. In summary, MCs would not be collapsing.

There are, however, several problems with such a classical view of molecular clouds being supported by turbulence against gravity. On the one hand, the studies for molecular clouds in the Solar Neighbourhood by different authors \citep{Lada+13, Lombardi+14, Zari+16, Lada+17, Pokhrel+21} show that star formation occurs fundamentally in the densest regions of molecular clouds, where the column density is of the order of $\SigmaMC\gtrsim$~500~$M_\odot$~pc\alamenos2. In this regime, the depletion time is about 10~Myr, 100 times faster than what typically assumed. Thus, the long depletion timescales of molecular gas at galactic scale cannot be taken as a proof of star formation being slow.

It should be noticed, furthermore, that large efficiencies cannot be taken as an argument against fast star formation, since the final star formation efficiency is reduced by: (i) the fact that the denser regions, which collapse and form stars faster than the rest of the cloud, contain only a small fraction of the total mass \citep{Ballesteros-Paredes+11b}; (ii) the fact that geometry factors do play a role \citep{Hartmann_Burkert07, Pon+12}, where filaments develop an efficiency $\sim10$~times lower than spheres \citep{Zavala-Molina+24}; and (iii) the large efficiency of stellar feedback (winds and UV radiation) in disrupting the parental cloud \citep[e.g., ][]{Kruijssen+18, Krause+20, Grudic+21, Kim+21}. Thus, the fact that CO clouds are mostly collapsing \citep{Hartmann+01} does not imply a large star formation rate.

Consistent with these short timescales is the observational evidence that molecular clouds have no stars older than 5~Myr \citep{Herbig78, Leisawitz+89, Ballesteros-Paredes+09b, Ballesteros-Paredes_Hartmann07, Kruijssen+18} implying that newborn stars manage to get rid of their parental molecular clouds after two free-fall times. { This result points} to a fast star formation scenario in which star formation is rapid and stellar feedback is efficient in destroying the parental cloud \citep[e.g., ][]{Ballesteros-Paredes+99b, Elmegreen00, Hartmann+01, Ballesteros-Paredes_Hartmann07, Bournaud+10, Colin+13, Hopkins+11, Roman-Zuniga+15, Kruijssen+18, Vazquez-Semadeni+19, Krause+20, Kim+21, Grudic+21}.

Regarding the virial state of molecular (and in particular, CO) clouds, there are several points to stress:

\begin{enumerate}
  \item While this is true that most of the CO clouds exhibit overvirial values, most of the mass is in the larger clouds. These exhibit values of $\alphavir\sim1-2$ \citep[e.g.,][]{Miville-Deschenes+17, Evans+21}. 

  \item Regarding the concerns on overestimating the gravitational energy due to simplifications \citep{Chevance+23}, we found the opposite effect: typically, self-gravity is underestimated. In fact, we found that collapsing clouds may exhibit overvirial values if their actual density structure is not properly taken into account \citep{Ballesteros-Paredes+18}.
  
  As a simple example, one can reconsider the derivation by \citet[][see above]{Chevance+23}. Since clumps and cores do have higher densities than their parent clouds, one should consider not only the scaling of the velocity dispersion but also that of the density, when estimating the virial parameter. For instance, assuming that the density profile scales as $\rho\propto l^{-\beta}$, then the virial parameter should scale as
\begin{equation}
  \alphavir \propto l^{\beta-1}.
  \label{eq:alphavir_vs_scale}
\end{equation}
As it can be seen, flat density profiles ($\beta < 1$) will satisfy the \citet{Chevance+23} condition that smaller structures are overvirial. But clumps and cores with density profiles steeper than $\beta>1$ will become more gravitationally bound than their parental cloud. Indeed, \citet[][see their Fig.~1]{Kauffmann+13} have shown that for different tracers, there are always clouds, clumps and cores with $\alphavir\simeq 2$ at all scales.

\end{enumerate}

{One of the problems with assuming turbulence as the physical agent supporting clouds against collapse is that turbulence, a highly dissipative phenomenon, must be driven continuously and homogeneously.} The apparent longevity of virial or pressure-confined supervirial clouds is, in a way, a wrong extrapolation from stellar cluster dynamics, where clusters are virial and long-lived because they are not dissipative.

We can analyze how a stellar system achieves virial balance to understand this misconception. Let us first assume super-virial initial conditions. The long-term evolution of such a system will be its expansion,  never reaching virial balance. If, on the contrary,  the stellar system has initial sub-virial conditions, it will collapse under its self-gravity and reach virial values after one free-fall time \citep{Noriega-Mendoza_Aguilar18}.

{In the case of molecular clouds the evolution is different: for a subvirial cloud, as collapse proceeds it will reach virial values as a consequence of the collapse itself \citep[e.g., ][]{Ballesteros-Paredes+18}. Thus, its kinetic motions are due to the collapse itself and not to any form of kinetic support. 

In the case of an initially super-virial cloud \citep[as it can be the case of a cloud being formed by the convergence of flows, see, e.g., ][]{Camacho+16, Camacho+23, Ballesteros-Paredes+18, Ganguly+22},} one cannot give for granted that the whole cloud will avoid collapse. {Statistically speaking, half of the motions will be converging \citep[a fact actually measured in numerical simulations, see][]{Vazquez-Semadeni+08, Camacho+16, Appel+23}, and will not produce support or expansion, but collapse \citep[see][]{Ballesteros-Paredes06}. Indeed, it should be recalled that, } in the current model of molecular cloud formation \citep{Ballesteros-Paredes+99a, Hennebelle_Perault99, Hartmann+01, Vazquez-Semadeni+07, Heitsch_Hartmann08, MacLow+17, Vazquez-Semadeni+19, Ibañez-Mejia+22}, clouds are formed by the confluence of overvirial diffuse gas streams\footnote{Note that these streams may have different origins, powered by winds or supernova, spiral arms, tidal forces from MC complexes, Toomre-type instabilities, etc., \citep[see, e.g., ][]{Dobbs+14, Padoan+16, Mao+20, Ramirez-Galeano+22, Ganguly+22}.}. The larger-than-two values of the virial parameter for molecular clouds do not contradict the result that clouds proceed to collapse as they are being formed. These streams collide and compress the gas, which cools down rapidly \citep{Hennebelle_Perault99} and becomes molecular \citep{Ballesteros-Paredes+99b, Hartmann+01, Bergin+04, Park+23}. Consequently, the denser parts collapse. The result after a fraction of the free-fall time is that the cloud starts exhibiting virial balance. Such balance, nonetheless, does not mean that the cloud is in equilibrium and will survive many free-fall times. The virial values of its energies only mean that the cloud is collapsing \citep[see, e.g., Figs.~8 and 3 in][respectively]{Vazquez-Semadeni+07, Ibañez-Mejia+22}. As we can see, overvirial clouds are part of the scenario of cloud formation, in which gravity takes over, and its own collapse leads up to the fundamental law of star formation, eq.~(\ref{eq:fundamental2}).

{ In summary, the appearance of overvirial CO clouds \citep[e.g.,][]{Evans+22}   is not in contradiction with the idea that most of the CO is collapsing gas: (a) the approximations performed in observational work may underestimate the actual gravitational content of molecular clouds \citep{Ballesteros-Paredes+18}; (b) even if overvirial parameters are real, at least half of the overvirial motions must be converging \citep[e.g., ][]{Vazquez-Semadeni+08, Camacho+16, Ballesteros-Paredes+18}, and thus, induce collapse; (c) at face value, most of the mass in molecular clouds at any level of he hierarchy exhibits $\alphavir\sim1-2$ \citep{Kauffmann+13}, and thus, they have been undergoing collapse \citep{Vazquez-Semadeni+07, Ballesteros-Paredes+11a, Noriega-Mendoza_Aguilar18}.}
}

\subsection{ The role of feedback in the KS relation}

\citet{Pokhrel+21} had argued that { magnetic supersonic turbulence produced by stellar feedback is an adequate candidate to explain the tight correlation within individual molecular clouds spanning several orders of magnitude in masses and SFR with constant $\eff$. { However, it should be noticed that the low values of $\eff$ estimated by these authors in local molecular clouds involve not only places where there is active star formation but also where there are no newborn stars \citep[see also][]{Zamora-Aviles+24}}}.

It should be stressed that in deriving the fundamental equation (\ref{eq:fundamental2}), we have not assumed the nature of the feedback and have shown that the constancy of $\eff$ is the very consequence of the gas that is involved in the collapse to form stars. Conversely, if turbulence dominates the star formation process, the natural timescale used in the fundamental equation should be the dynamical crossing time $\taudyn$, rather than the free-fall time. In reality, the relevance of the stellar feedback is not to { support clouds over many free-fall times,  maintaining low the} `efficiency per free-fall' time but to shut down the star formation process \citep{Hartmann+01, Kruijssen+18, Grudic+21}. 

{
It is important to mention that the fact that} the mass in eq.~(\ref{eq:eff:definition}) is the mass involved in the process of collapse of the cloud does {not} imply that all that mass {will} end up forming stars within one free fall time with a huge star formation rate. The reason is that MCs are highly inhomogeneous and fractal \citep[e.g.,][]{Falgarone+91}, and then, the free-fall timescales within clouds have substantial variations from place to  place. { Then, although most of the mass of a molecular cloud is at $n\sim100$~cm\alamenos3, dense cores ($n\gtrsim10^5$~cm\alamenos3) containing only a small fraction of the total mass of the cloud, will form stars in a much shorter timescale, initiating the termination of the star formation process and dispersing the otherwise collapsing cloud well before the less-dense regions {significantly develop the} collapse.}

\subsection{Clouds with lognormal PDFs cannot have a constant `efficiency per free-fall time'}\label{sec:not:lognormal}

The results from section \ref{sec:SigmaSFR-Sigmagas2:new} have an important implication for models of molecular cloud dynamics. As we have shown, $\eff$ is constant as long as the power-law index of the column density and the power-law index of the KS relation follows eq.~(\ref{eq:KS-PDF:relation}). In more general terms, it can be demonstrated that 
\begin{equation}
  \eff \propto \frac{M^{N-3/2}}{L^{2N-7/2}}
  \label{eq:eff:pdf}
\end{equation}
which becomes constant if $\alpha_1$ in  eq.~(\ref{eq:alphauno}), and $\alpha_2$ in eq.~(\ref{eq:alphados}), are equal, as we showed in \S\ref{sec:SigmaSFR-Sigmagas2:new}. But strictly speaking, the mass and the size are given respectively by eqs.~(\ref{eq:mass:pdf}) and the square root of (\ref{eq:surface:pdf}). If the column density PDF of a cloud is a lognormal function, the `efficiency per free-fall time', eq.~(\ref{eq:eff:pdf}), cannot be a constant but a quantity that will depend on the column density of the cloud. Thus, clouds with lognormal column density PDFs cannot have a constant `efficiency per free-fall time'. In a sense, this is another demonstration that turbulence cannot be responsible for keeping $\eff$ constant, since turbulence-regulated clouds exhibit lognormal column density PDFs. 

\subsection{Some comments about the `efficiency per free-fall time'}\label{sec:eff}

Strictly speaking, the so-called `efficiency-per-free-fall-time' is the ratio between the free-fall time $\tauff$ and the depletion time, i.e., the time in which the cloud will convert all its mass into stars, 
\begin{equation}
  \tau_{\rm depl} = \bigg( \frac{\Mtotstarstausf/\tausf}{\Mgas} \bigg)^{-1} .
  \label{eq:depletiontime}
\end{equation}
There are some caveats, however, with this name and definition. On the one hand, it is implicit in its name that clouds could last many free-fall times. Indeed, if clouds do not last long, there will not be a need to define an efficiency other than the final efficiency of star formation. In addition, it also suggests that this quantity should remain constant over those many free-fall times, although there is no warranty that this will be the case. Finally, there are some problems with the observational estimations of this quantity, { at least in the case of local molecular cloud observations}. For instance, to compute $\eff$ for local clouds, one needs to know the timescales $\tausf$ of the different evolutionary stages of protostars (Class 0, I, and II). Those timescales have been estimated by assuming stationarity, i.e., that the ratio of objects in two different stages is proportional to the ratio of timescales in those stages, i.e., 
\begin{equation}
  \frac{N_1}{N_2} = \frac{\tau_1}{\tau_2}
\end{equation}
\citep[see, e.g., ][]{Lee_Myers99, Ward-Thompson+07}. However, { star formation may not be stationary}. For instance, the embedded phase of a protostar may depend on its proximity to a massive star, and thus, there is no guarantee that the timescale in a particular stage is always the same for all objects. This will affect the estimations of the depletion time and, thus, the estimations of the `efficiency per free-fall time'.

{ Finally, the very purpose of defining efficiencies is to estimate, given a physical process, how much of a certain physical property can be obtained from an original pool. By definition, this quantity has to be smaller than one. Thus, the efficiency of star formation is, then, the final mass of stars compared to the total mass of gas that was part of the cloud during the process of star formation. However, in the case of the `efficiency per free-fall time', $\eff$, there can be a situation in which $\eff$ is larger than unity \citep[see, e.g., Fig.~7 of][]{Clark_Glover14}, just because the rate of star formation at a given time may be larger than the rate of collapse of the cloud, $\Mdense/\tauff$. Thus, we argue that the `efficiency per free-fall time' is a misleading name for the parameter $\eff$, and it should be called simply the ratio between the star formation rate and the gas infall rate. In fact, naming it in this way will allow a better understanding of its meaning. For instance, a constant $\eff$ will imply that the star formation rate changes at the same rate as the infall rate, basically because both of them are deeply related, and because the latter is the cause of the former. 
}

\subsection{Collapsing clouds also exhibit low values of `efficiency per free-fall time'}

It is well known that a collapsing cloud may have large values of the final star formation efficiency if it has no stellar feedback. Interestingly, using numerical simulations, \citet{Zamora-Aviles+24} show that collapsing clouds exhibit low values of $\eff$ ($\sim0.03$), compatible with the inferred observational estimations \citep[e.g., ][]{Krumholz+19, Pokhrel+21}. These authors also show that $\eff$ remains constant at different levels of the hierarchy of the cloud, \citep[just as the clouds in the Solar Neighbrohood,][]{Lombardi+14, Zari+16, Pokhrel+21}, and that it remains constant in time, confirming our analytic result that collapsing clouds must exhibit a constant $\eff$.

These results are at odds with the intuition. This apparent inconsistency has been explained by \citet{Vazquez-Semadeni+19} and \citet{Bonilla-Barroso+22} as a consequence of how estimations of $\eff$ are performed. Indeed, the `efficiency per free-fall time' contains the mass in gas and its free-fall timescale. Both quantities are computed from observations at the present day. But suppose clouds form from the diffuse medium and become gravitationally unstable as they reach larger densities and smaller temperatures \citep{Hartmann+01}. In that case, the current free-fall time is not representative, as it changes every instant. As pointed out by \citet{Vazquez-Semadeni+19} and \citet{Bonilla-Barroso+22}, the current free-fall timescale of cores severely underestimates the time lapse that a larger and less dense cloud has to spend to achieve its present state. 

{
\section{Gravity or turbulence: What is new?}\label{sec:whatsnew}

In the framework of the collapse scenario \citep[][ and references therein]{Hartmann+01, Ballesteros-Paredes+11a, Elmegreen18, Vazquez-Semadeni+19, Ibañez-Mejia+22}, it is natural to understand eq.~(\ref{eq:fundamental2}) as the fundamental equation of star formation because, indeed, it is hard to form a molecular cloud from the diffuse ISM that can last several free-fall times without collapsing into different centres, forming stars and dispersing. The idea that this equation is fundamental has already been proposed before. For instance, \citet{Krumholz+12} show that this correlation explains the observational data (mainly extragalactic) substantially better. In addition, \citet{Clark_Glover14} have found that any region that is cold and dense enough to collapse follows eq.~(\ref{eq:fundamental2}). These authors showed that the molecules are only a proxy for the cold and dense regions that collapse because they tend to be dissociated at larger temperatures or lower densities. 

{ There are, however, important differences between our approach and previous work favouring the fundamental equation (\ref{eq:fundamental2}). For instance, } the small values of $\eff$, have led different authors to argue that turbulence must keep clouds forming stars at a low pace over many free-fall times. This idea has been taken to the extreme to consider that star formation is a slow process \citep[see also][]{Krumholz_McKee05, Krumholz_Tan07, Krumholz+19, Evans+22}. In contrast, our interpretation is that, being dense and cold, molecular clouds collapse, form stars, and disperse rapidly. There are either observational and theoretical evidences of this \citep{Herbig78, Leisawitz+89, Ballesteros-Paredes+99b, Ballesteros-Paredes_Hartmann07, Heitsch_Hartmann08, Kruijssen+18, Vazquez-Semadeni+19, Kim+21}.

{
Another relevant difference with \citet{Krumholz+12} is between their approach and ours.} For us, all star-forming clouds (most of them molecular) are at similar densities ($n\sim 100$~cm\alamenos3). That is precisely the reason for the linear correlations between $\Sigmasfr$ and $\SigmaMC$ observed by different authors in CO  \citep[e.g., ][]{Bigiel+08, Sun+23, Ellison+24}. For them, observations in CO may trace different densities, according to  $\rhogas = \Sigmagas/h_z$, and thus, different free-fall times. To give an example, from eqs.~(6) and (8) in \citet{Krumholz+12}, one can obtain volumetric densities\footnote{We used characteristic values of Milky Way-type galaxies (e.g., orbital frequency $\Omega\sim$~10\alamenos{15}~sec\alamenos1; Toomre parameter $Q\sim1$, and column densities of the galaxy $\Sigmagas\sim10-100$~\Msun~pc\alamenos2).} of $n\sim1$ and $\sim20$~cm\alamenos3 respectively, substantially smaller than the characteristic volumetric densities of molecular clouds.

By assuming the density of MCs to be the characteristic density from which most of the CO emission comes, our estimation of the free-fall time is local. In contrast, \citet{Krumholz+12} are deriving local quantities (density, free-fall time) from global ones, as the column density and the Toomre parameter at a given place in the galaxy. This approach would be valid if the free-fall time of molecular clouds were comparable to the crossing time of the galaxy in those places. But as we pointed out in \S\ref{sec:pressure}, this is not the case. For instance, the vertical crossing timescale $h_z/\sigma_z$ (with $h_z$ and $\sigma_z$ the scale height and velocity dispersion in the vertical direction) is at least 10 times larger than the free-fall time of a cloud at $n\sim100-400$~cm\alamenos3. As for the Toomre instability, the crossing time is even longer since the distances to travel at similar velocities are even larger.

It should be noted that if the Toomre instability played a role in the collapse of molecular clouds, the gravitational energy contribution from the stellar disk would be comparable to the gravitational energy of the cloud itself. But the contribution from the stellar disk to the gravitational budget of molecular clouds is negligible \citep[e.g., ][]{Suarez+12, Ramirez-Galeano+22}. The stellar disk matters in terms of increasing the background density of the atomic gas from where molecular clouds would be formed, allowing fast MC formation \citep{Ramirez-Galeano+22}.

Our work, however, resembles that of \citet{Elmegreen18}, who tried to understand the different KS correlations in terms of the same fundamental law. In principle, both works agree that clouds must be collapsing. There are also some differences between our work and his, which should be clarified. In particular,

\begin{enumerate}

  \item \citet{Elmegreen18} argues that all gas is collapsing, but at different rates. In contrast, we argue that eq.~(\ref{eq:fundamental2}) has to be considered basically for collapsing gas only. When including gas that is not collapsing (or collapsing at substantially different rates), the linearity of eq.~(\ref{eq:fundamental2}) is necessarily broken, and the power-law index of the KS relation would depend on how the HI/H$_2$ ratio varies in each galaxy.

  \item We argue that the linear relation is valid in all its forms, but care has to be taken when different comparisons are made. For instance, for us, the linear correlation between the star formation rate and the gas mass \citep{Gao-Solomon04, Wu+05, Lada+10} is a direct application of eq.~(\ref{eq:fundamental2}), provided that the observed regions in each work have similar mean densities. However, \citet{Elmegreen18} argues that it may be a consequence of observational artefacts. 
  
  \item Our approach explaining the linear SF law for molecular gas in resolved galaxies \citep[relation KS-1a in][]{Elmegreen18} is basically the same: cold dense gas collapses within 1--2~free-fall timescales. Nonetheless, we argue that there is no need to consider total gas to understand this relation since HI is substantially less dense than CO. Thus, its process of collapse is most likely to be disrupted by the star formation event produced by the cold, dense gas, long before {it significantly develops the collapse.}

  \item As for the non-linear correlations, \citet{Elmegreen18} { distinguishes} between the KS-1.5 and the KS-2 relations (the power laws with power-law indexes of 1.5 and 2). So, in order to explain the KS-1.5 relation \citep[i.e., the original result by ][ with a power-law index of 1.5]{Kennicutt98} for whole galaxies, \citet[][\S~2.2]{Elmegreen18} invokes the density $\rho = \Sigmagas/2H$, with $H$ the scale-height of the disk. And since $\tauff\propto \rho^{-0.5}$, if the scale height is constant through the galaxy or between galaxies, then the KS relation with power-law 1.5 follows naturally. We believe that there is no need to invoke the constancy of the scale height of disk galaxies. Especially, knowing that the gas disks are flared \citep[e.g., ][]{Kalberla_Kerp09, Yim+14} and that there is not a single universal scale-height of the gaseous disk galaxies \citep{ManceraPiña+22}. Our proposal is, instead, that galaxies with larger column densities are more likely to have a larger fraction of dense, cold (molecular) gas \citep[e.g., ][]{Park+23}, which will collapse faster. Thus, the increase in total column density implies a superlinear increase of such a dense, cold gas, which implies a superlinear increase in star formation rate.

  \item With the previous interpretation of the relative abundance of dense {\it vs.} non-dense gas, we try to explain not only these galaxies with index 1.5 but also the variety of non-linear correlations with power-law indexes between 1.5 and 2.3 found when total gas is analyzed \citep[e.g., ][]{Bigiel+08}. We argue that the actual increase of $\Sigmasfr$ with total $\Sigmagas$ depends upon the local conditions of extinction, UV radiation and abundances that will allow (or not) the gas to cool down and become dense enough to collapse faster \citep[see][]{Clark_Glover14}. In contrast, \citet[][see \S~2.4]{Elmegreen18} focus on the power-law index of 2 in terms of pressure equilibrium provided mainly by the self-gravity of the disk.
  
  \item We also show that the internal structure of molecular clouds in intimately related to the KS relation they exhibit, and thus, that the power-law index of the column density PDF that each cloud has is intimately related to the power-law index of its KS relation.

  \item \citet{Elmegreen18} shows that a cloud with a density profile of a single isothermal sphere in virial equilibrium does follow the quadratic KS relation $\Sigmasfr\propto \SigmaMC^2$, a result found by \citet{Pokhrel+21}. Indeed, it is easy to demonstrate that a cloud with such a profile will have a free-fall time $\tauff\propto\SigmaMC^{-1}$, producing the quadratic correlation. And although the virialization of the cloud is not needed for these clouds to follow the KS relation, it should be recalled that virial balance is the natural outcome of collapse \citep{Vazquez-Semadeni+07, Noriega-Mendoza_Aguilar18, Ballesteros-Paredes+20}.

\end{enumerate}

There are other differences between our work and that of \citet{Elmegreen18}, being the main ones (a) our interpretation of the Silk-Elmegreen and Pressure-regulated correlations as basically the same fundamental law disguised by different factors; (b) our deduction that the column density PDF of individual molecular clouds is intimately related to their KS relation, as \citet{Lada+10} conjectured.  

}

\section{Conclusions}\label{sec:conclusions}

We started this project aimed at understanding the variety of Kennicut-Schmidt correlations in the literature,  since not all seemed compatible or consistent with each other. To do so, we first derived the {relation} 
\begin{equation}
  \promedio{\mathrm{SFR}}_\tausf = \eff\ \bigg(\frac{\Mcollapsing}{\tauff}\bigg).
\end{equation}
Since this was done from the pure definition of star formation rate and basic algebra, we call it the fundamental law of star formation. We then demonstrate that if the `efficiency per free-fall time' is constant, { it} is because the mass under consideration is in the process of Jeans fragmentation and, thus, must be collapsing.

We further have shown that the different correlations observed in the literature, galactic or extragalactic, can be inferred from the fundamental law, provided there is an understanding of the object's structure in question. The first one, $\Sigmasfr\propto \Sigmagas/\tauff$, shown by \citet{Pokhrel+21}, is just a direct application of the law. The following two, namely $\Nysomath-\Mdense$ \citep{Lada+10} and the linear correlation between the infrared luminosity as a tracer of star formation rate, and the HCN luminosity as a tracer of the mass in dense gas $L_{\rm IR}-L_{\rm HCN}$ \citet{} \citep{Gao-Solomon04}, can also be understood in terms of the fundamental equation of star formation (\ref{eq:fundamental}) by noticing that the objects under consideration in each correlation have the same density regime and, thus, have similar free-fall times. But there is a third type of correlation, more similar to the \citet{Kennicutt98} correlation,  $\Sigmasfr\propto\Sigmagas^N$, which, in order to be compatible with the fundamental law, requires an understanding of the internal structure of the cloud, namely, the column density PDF of the cloud, and its mass-size relation at different levels of the hierarchy of the cloud.

We then moved to the extragalactic correlations. We found that these could also be explained through the fundamental equation of star formation. In particular, we interpret the superlinear behaviour of the Kennicutt-Schmidt relationship as a consequence of { the correlation between the molecular gas abundance and the column density of gas (HI+H$_2$)  \citep[see also][]{Park+23} caused by the increasing self-shielding of gas against the dissociating UV radiation, as column density increases}. The remaining correlations were explained in terms of the different methodologies or assumptions without the need to explain further physical processes. 

It should be noticed, furthermore, that the observed constancy of $\eff$ implies that star formation is regulated by the collapse of the cloud only. The role of galactic rotation is not to regulate the star formation but to regulate the formation of HI clouds and, thus, only indirectly, to regulate the formation of molecular clouds. { On the other hand, the role of stellar feedback is to shut down the local star formation event \citep[e.g., ][]{Bournaud+10, Colin+13, Kim+21, Hopkins+11}, and thus, to keep low the global star formation efficiency.} But the rate at which molecular gas is transformed into stars is only regulated by the gravitational collapse itself, {and the fact that $\eff$ is found to be constant implies that the ratio of the star formation rate over the gas infall rate is the same during the star formation process.}

\section*{Data Availability}

The data underlying this paper will be shared on a reasonable request to the corresponding author.

\section*{Acknowledgments}

We thank an anonymous referee for helping us improve this work's quality.
J.B-P.  acknowledges UNAM-DGAPA-PAPIIT support through grant number IN-114422. He also 
acknowledges Paris-Saclay University’s Institute Pascal 'Interstellar Institute’ meeting in 2023, in which invaluable discussions with the participants led to developing the idea behind this work.
M.Z.A. acknowledges support from CONAHCYT grant number 320772. 
C.R.Z. acknowledges support from programs UNAM-DGAPA-PAPIIT IG101723 and CONAHCYT CB2018-A1-S-9754. 
A.P. acknowledges financial support from the UNAM-DGAPA-PAPIIT IG100223 grant and the Sistema Nacional de Investigadores of CONAHCyT, M\'exico.
B.C.S. acknowledges financial support from the UNAM-DGAPA-PAPIIT IN108323 grant.
K.G.D. acknowledges a scholarship from CONAHCyT.
V.C. acknowledges a postdoctoral fellowship from CONAHCyT.
E.J.A acknowledges support from UNAM-PAPIIT project IA-102023, and from CONAHCyT Ciencia de Frontera project ID CF-2023-I-506.´´ 
A.G. acknowledges support from UNAM-PAPIIT project IN115623.
J.B.P, C.R.Z, and A.P. acknowledge support from CONAHCYT grant number 86372.

\bibliographystyle{mnras}
\bibliography{MisReferencias} 

\appendix





\bsp	
\label{lastpage}
\end{document}

%% file: definitions.tex
\def\apj{ApJ}
\def\apjl{ApJL}
\def\aap{A\& A}

\def\araa{ARA\&A}
\def\mnras{MNRAS}

\def\aj{{AJ}}
\def\apjs{ApJS}

\def\mnras{{MNRAS}}
\def\pasj{{PASJ}}
\def\pasp{{PASP}}

\def\ala#1{$^{#1}$}
\def\alamenos#1{$^{-#1}$}

\newcommand{\alphauno}{{\alpha_1}}
\newcommand{\alphados}{{\alpha_2}}

\newcommand{\alphavir}{\alpha_{\rm vir}}

\newcommand{\CR}{\rm CR}    

\def\diezala#1{$10^{#1}$}
\def\diezalamenos#1{$10^{-#1}$}

\newcommand{\deltavl}{\delta_{v_l}}

\newcommand{\eff}{{\epsilon_\mathrm{ff}}}    %
\newcommand{\eorb}{{\epsilon_\mathrm{orb}}}    %

\newcommand{\Mjeans}{M_{\rm Jeans}}    %

\newcommand{\Mcollapsing}{M_\mathrm{collapse}}    %
\newcommand{\Mdense}{M_\mathrm{dense~gas}}    %

\newcommand{\Mgas}{M_\mathrm{gas}}    %

\newcommand{\Mtotstarstausf}{{M_\mathrm{*,\tausf}}}

\newcommand{\Msun}{$M_\odot$}    %
\newcommand{\Njeans}{N_{\rm Jeans}}    %

\newcommand{\Ntotstarstausf}{{N_\mathrm{*,\tausf}}}

\newcommand{\Nysomath}{{N_{\rm YSO}}}

\def\promedio#1{\langle{#1}\rangle}

\newcommand{\PDE}{P_{\rm DE}}

\newcommand{\rhogas}{\rho_\mathrm{gas}}
\newcommand{\rhosfr}{\rho_\mathrm{SFR}}

\newcommand{\SFR}{{\rm SFR}}

\newcommand{\Sigmagas}{\Sigma_{\rm gas}}
\newcommand{\Sigmagascero}{\Sigma_{\rm gas,0}}
\newcommand{\SigmaMC}{\Sigma_{\rm mol}}
\newcommand{\Sigmasfr}{\Sigma_{\rm {SFR}}}

\newcommand{\sigmagasz}{\sigma_{{\rm gas,}z}}

\newcommand{\taudepl}{\tau_{\rm depl}}
\newcommand{\taudyn}{\tau_{\rm dyn}}
\newcommand{\tauff}{{\tau_{\rm ff}}}
\newcommand{\tauorb}{{\tau_{\rm orb}}}
\newcommand{\tausf}{{\tau_{\mathrm{YSO}}}}
\newcommand{\tauyso}{{\tau_{\mathrm{YSO}}}}